\documentclass[remotesensing,article,accept,pdftex,moreauthors]{Definitions/mdpi} 
\usepackage{tikz,geometry,url}
\usetikzlibrary{positioning,arrows.meta,shapes.geometric}
\usepackage{adjustbox}
\newlength\mylength

\newcommand\aj{Astron.~J.}%
\newcommand\apjl{Astrophys.~J.~Lett.}%
\newcommand\apj{Astrophys.~J.}%
\newcommand\grl{Geophys.~Res.~Lett.}%
\newcommand\jgr{J.~Geophys.~Res.}%
\newcommand\jgrs{J.~Geophys.~Res.~Supp.}%
\newcommand\icarus{Icarus}%
\newcommand\nat{Nature}%
\newcommand\planss{Planet.~Space~Sci.}%
\newcommand\ssr{Space~Sci.~Rev.}%
\newcommand\psj{Planet.~Sci.~J.}%
\newcommand\areps{Ann.~Rev.~Earth~Planet.~Sci.}%
\firstpage{1} 
\pubvolume{1}
\issuenum{1}
\articlenumber{0}
\pubyear{2022}
\copyrightyear{2022}
\externaleditor{{Academic Editors:} Imke de Pater and Yamila Miguel} % please add Academic Editor information if possible
\datereceived{31 October 2022} 
\dateaccepted{20 November 2022} 
\datepublished{} 
\hreflink{https://doi.org/} % If needed use \linebreak
\Title{Giant Planet Observations in NASA's Planetary Data System}

\TitleCitation{Giant Planet Observations in NASA's Planetary Data System}

\Author{Nancy J. Chanover
 $^{1,*}$\orcidA{}, James M. Bauer $^2$\orcidB{}, John J. Blalock $^3$\orcidG{}, Mitchell K. Gordon $^4$\orcidD{}, Lyle F. Huber $^1$\orcidH{}, \linebreak Mia J. T. Mace $^4$\orcidF{}, Lynn D. V. Neakrase $^1$\orcidI{}, Matthew S. Tiscareno $^4$\orcidE{} and Raymond J. Walker $^{5}$\orcidC{}}

\AuthorNames{Nancy J. Chanover, James M. Bauer, John J. Blalock, Mitchell K. Gordon, Lyle F. Huber, Mia J. T. Mace, Lynn D. V. Neakrase, Matthew S. Tiscareno and Ray J. Walker}

\AuthorCitation{Chanover, N.J.; Bauer, J.M.; Blalock J.J.; Gordon, M.K.; Huber, L.F.; Mace, M.J.T.; Neakrase, L.D.V.; Tiscareno, M.S.; Walker, R.J.}
\address{%
$^{1}$ \quad Astronomy Department, New Mexico State University, Las Cruces, NM 88003, USA; \linebreak lhuber@nmsu.edu (L.F.H.); lneakras@nmsu.edu (L.D.V.N.)
\\
$^{2}$ \quad Department of Astronomy, University of Maryland, College Park, MD 20742, USA; gerbsb@umd.edu\\
$^{3}$ \quad U.S. Geological Survey, Astrogeology Science Center, Flagstaff, AZ 86001, USA; jblalock@usgs.gov\\
$^{4}$ \quad Carl Sagan Center, SETI Institute, 339 Bernardo Avenue \#200, Mountain View, CA 94043, USA; mgordon@seti.org (M.K.G.); mmace@seti.org (M.J.T.M.); mtiscareno@seti.org (M.S.T.)\\
$^{5}$ \quad Department of Earth, Planetary %MDPI: Is this necessary? Can it be removed?. %%NJC: removed comma
 and Space Sciences, University of California, Los Angeles, CA 90095, USA; rwalker@igpp.ucla.edu}

\corres{Correspondence: nchanove@nmsu.edu; Tel.:  +1-575-646-2567}

\abstract{%A single paragraph of about 200 words maximum. For research articles, abstracts should give a pertinent overview of the work. We strongly encourage authors to use the following style of structured abstracts, but without headings: (1) Background: place the question addressed in a broad context and highlight the purpose of the study; (2) Methods: describe briefly the main methods or treatments applied; (3) Results: summarize the article's main findings; (4) Conclusion: indicate the main conclusions or interpretations. The abstract should be an objective representation of the article, it must not contain results which are not presented and substantiated in the main text and should not exaggerate the main conclusions.}
While there have been far fewer missions to the outer Solar System than to the inner Solar System, spacecraft destined for the giant planets have conducted a wide range of fundamental investigations, returning data that continues to reshape our understanding of these complex systems, sometimes decades after the data were acquired. These data are preserved and accessible from national and international planetary science archives. For all NASA planetary missions and instruments the data are available from the science discipline nodes of the NASA Planetary Data System (PDS). Looking ahead, the PDS will be the primary repository for giant planets data from several upcoming missions and derived datasets, as well as supporting research conducted to aid in the interpretation of the remotely sensed giant planets data already archived in the PDS.}

\begin{document}

%%%%%%%%%%%%%%%%%%%%%%%%%%%%%%%%%%%%%%%%%%
%\setcounter{section}{-1} %% Remove this when starting to work on the template.
%\section{How to Use this Template}

%The template details the sections that can be used in a manuscript. Note that the order and names of article sections may differ from the requirements of the journal (e.g., the positioning of the Materials and Methods section). Please check the instructions on the authors' page of the journal to verify the correct order and names. For any questions, please contact the editorial office of the journal or support@mdpi.com. For LaTeX-related questions please contact latex@mdpi.com.%\endnote{This is an endnote.} % To use endnotes, please un-comment \printendnotes below (before References). Only journal Laws uses  (\highlighting.

% The order of the section titles is: Introduction, Materials and Methods, Results, Discussion, Conclusions for these journals: aerospace,algorithms,antibodies,antioxidants,atmosphere,axioms,biomedicines,carbon,crystals,designs,diagnostics,environments,fermentation,fluids,forests,fractalfract,informatics,information,inventions,jfmk,jrfm,lubricants,neonatalscreening,neuroglia,particles,pharmaceutics,polymers,processes,technologies,viruses,vision

%%%%%%%%%%%%%%%%%%%%%%%%%%%%%%%%%%%%%%%%%%%%%
\section{Introduction}\label{sec:Intro}

%The introduction should briefly place the study in a broad context and highlight why it is important. It should define the purpose of the work and its significance. The current state of the research field should be reviewed carefully and key publications cited. Please highlight controversial and diverging hypotheses when necessary. Finally, briefly mention the main aim of the work and highlight the principal conclusions. As far as possible, please keep the introduction comprehensible to scientists outside your particular field of research. Citing a journal paper \cite{ref-journal}. Now citing a book reference \cite{ref-book1,ref-book2} or other reference types \cite{ref-unpublish,ref-communication,ref-proceeding}. Please use the command \citep{ref-thesis,ref-url} for the following MDPI journals, which use author--date citation: Administrative Sciences, Arts, Econometrics, Economies, Genealogy, Humanities, IJFS, Journal of Intelligence, Journalism and Media, JRFM, Languages, Laws, Religions, Risks, Social Sciences, Literature.

% big picture motivation:
Spacecraft exploration of the outer Solar System began with the arrival of Pioneer~10 at Jupiter in 1973, and continues to the present day with the Juno spacecraft in orbit around Jupiter. Unlike the more frequent orbiter and lander missions to destinations in closer proximity to Earth ({e.g.}, the Moon or Mars), missions to the outer Solar System have been more scarce due to cost, power, and travel time constraints. Nevertheless, spacecraft destined for the giant planets have conducted a wide range of fundamental investigations, returning data that continues to reshape our understanding of these complex systems, sometimes decades after the data were acquired.

%PDS history paragraph:
Early in the history of NASA's planetary missions, data archiving was somewhat unregulated: data were typically delivered from missions to Principal Investigators (PIs), who were individually responsible to clean, process, and document the data, and then to deposit them with the National Space Science Data Center Archive (NSSDCA) for permanent storage and public access. However, there was a realization in the late 1980s that careful stewardship of data was required to ensure both %MDPI: Please confirm if the italics is unnecessary and can be removed. The following highlights are the same.
%% NC: italics were unnecessary and were removed
its accessibility and its usability for future generations of researchers.  Not only were the data degrading (whether stored on analog media such as photographic film or scan tapes, or on digital media such as magnetic tapes), but subsequent attempts by others to use the data were failing despite conscientious efforts on the part of the PIs to provide sufficient documentation. It was recognized that these remote sensing (and also \textit{in situ}) %MDPI: We removed the italics. Please confirm this revision. 
%% NC: we do not agree with this revision; as in situ is a Latin term, it should remain italicized
data represent a national treasure and a significant investment that must be preserved and maintained indefinitely.

NASA's Planetary Data System (PDS) was established with this aim in mind, following recommendations made by the Committee on Data Management and Computation (CODMAC) on behalf of the Space Science Board, National Academy of Sciences  \citep{Bernstein1982, Arvidson1986}, with the PDS being one of a series of data systems recommended (the others covering other scientific disciplines such as Heliophysics and Astrophysics). CODMAC reviewed several archiving approaches and concluded that planetary datasets should reside at research institutions associated with the relevant scientific community, ensuring that scientists with direct experience in working with the data are closely involved in the archiving and curation process as well as the development of tools for data search, analysis, visualization, and planning observations. 

% {MM
% sources: 
% - PDS Decadal White Paper https://pds-ppi.igpp.ucla.edu/AGU_2021/PDS_Decadal_White_Paper.pdf
% - 2017 Roadmap Study https://pds.nasa.gov/home/about/PlanetaryDataSystemRMS17-26_20jun17.pdf
%PDE report from last year https://science.nasa.gov/science-pink/s3fs-public/atoms/files/PDE%20IRB%20Final%20Report.pdf
%}

\begin{figure}[H]
\includegraphics[width=13.5 cm]{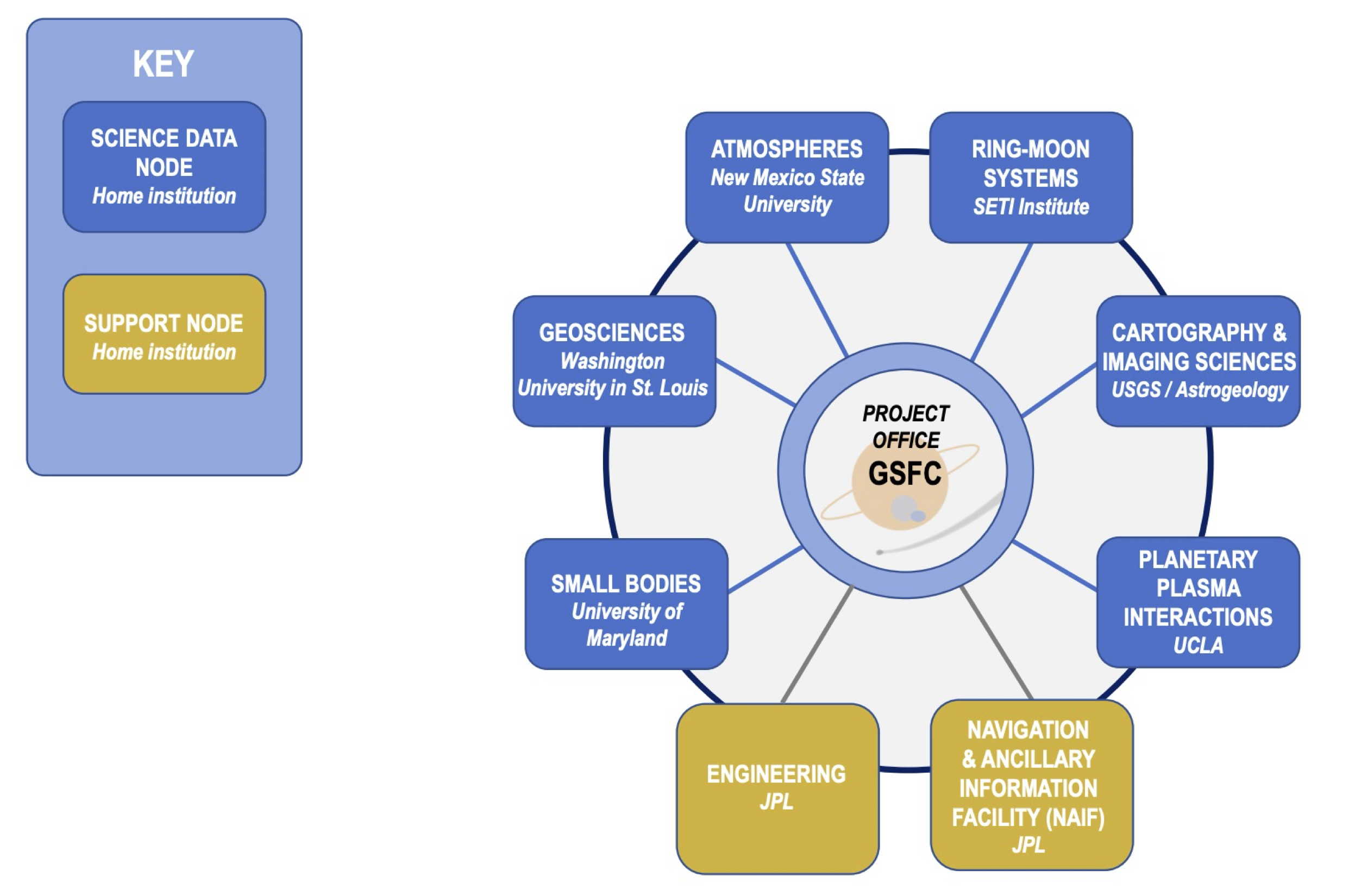}
\caption{Organizational structure of the PDS. The Discipline Nodes are shown in blue boxes. The support nodes, shown in gold boxes, contain experts in areas that cover numerous missions. The Radio Science Subnode, currently at RMS, supports data providers and users with the specialized needs of radio science data.
% NC: let us omit this; not relevant for this paper
%SBN is a distributed node, consisting of  the Comet Subnode (University of Maryland College, College Park, MD),  the Asteroid/Dust Subnode (Planetary Science Institute, Tucson, AZ), and the Minor Planet Center (Harvard-Smithsonian Center for Astrophysics, Cambridge, MA).
\label{fig:orgchart}}
\end{figure}

% PDS organization description:
To that end, the PDS is structured as a federated system of six Discipline Nodes (DNs) that provide unique expertise in various sub-disciplines of planetary science (Figure~\ref{fig:orgchart}) \cite{PDS2017}. Of these DNs, all but the Geosciences Node archive key remote sensing data sets related to the giant planets (specific examples are given in Section %MDPI: We changed \S to Section. Please confirm this revision.
%% NJC: this revision is fine
 \ref{sec:Data}): 
\begin{itemize}
	%https://pds.nasa.gov/home/about/node-descriptions.shtml
	\item \textls[-15]{The Atmospheres Node %MDPI: 1. Footernote is not allowed in this journal we have move it to the maintext in the brackets, please confirm. 2. Please add the access date (format: Date Month Year), e.g., accessed on 1 January 2020. the same as all below.
%% NJC: OK, done.
	 (ATM, {\url{https://pds-atmospheres.nmsu.edu}}, accessed on 30 October 2022) archives all non-imaging atmospheric data from planetary missions (excluding Earth observations),  ground-based observations, and planetary analog, laboratory and field measurements.}
	 \item The Cartography and Imaging Sciences Node (IMG, {\url{https://pds-imaging.jpl.nasa.gov}}, accessed on 30 October 2022) archives digital image collections from planetary missions, and supports cartographic and geospatial data analysis.
	 \item The Geosciences Node (GEO, {\url{https://pds-geosciences.wustl.edu}}, accessed on 30 October 2022) archives and distributes digital data related to the study of the surfaces and interiors of terrestrial planetary bodies. 
	 \item The Planetary Plasma Interactions Node (PPI, {\url{https://pds-ppi.igpp.ucla.edu}}, accessed on 30 October 2022) archives data related to the study of the interaction between the solar wind and planetary winds with planetary magnetospheres, ionospheres and surfaces.
	 \item The Ring-Moon Systems Node (RMS, {\url{https://pds-rings.seti.org}}, accessed on 30 October 2022) archives data relevant to outer planetary systems, with a focus on individual data products within their original context. This includes remote sensing data (images, imaging spectrometer, and occultations) for systems beyond the asteroid belt (that is, Jupiter through Pluto). RMS also hosts the Radio Science Sub-Node (RSSN), which assists all of PDS with the ingestion and curation of radio science data, including gravity science and~occultations.
	 \item The Small Bodies Node (SBN, {\url{https://pds-smallbodies.astro.umd.edu}}, accessed on 30 October 2022) archives mission, ground-based, and laboratory data for objects generally described as comets, asteroids and interplanetary dust. This includes dwarf planets, objects in the Kuiper Belt and the Oort cloud, Centaurs, and small planetary satellites, as well as observations of the giant planets acquired en route to small body targets. 
\end{itemize}

Data standards, {i.e.}, the technical specifications used to describe data storage, sharing, and interpretation, are crucial for information transfer and interoperability. PDS standards ensure consistent description of planetary science data so that data providers, programmers, and end-users all know what to expect when creating and working with PDS files \cite{PDS4concepts}.  Historically, the PDS Requirements Review \cite{Arvidson1986} laid out an ambitious design for preserving, cataloging, and distributing planetary datasets to users, all during a time that predated the widespread use of the internet. It consisted of a proposed three-stage rollout:
	\begin{itemize}
		\item PDS version 1.0: Focused on software infrastructure and catalog design (1990)
		\item PDS version 2.0: Basic operations in the central cataloging and distribution node, with two selected DNs being brought up to full operational status as test cases (1992--1993)
		\item PDS version 3.0: ``Final'' release with all DNs fully operational and user support in place (1994--1995)
\end{itemize}

By the mid-1990s, following feedback from users and data preparers, two major changes were implemented, resulting in what came to be known as PDS3 data standards (referring to v3.0 in the list above). First, major simplifications were made to the catalog file structures in order to streamline the process for users to locate and order datasets, and second, an attempt was made to rationalize and standardize the keywords used in labels by imposing some minimal metadata requirements. %Currently, all data are in the process of being migrated to PDS4, which imposes stricter standards,  to ensure better preservation, standardization, and automation}. 

Fifteen years following the introduction of PDS3, the process of developing and implementing new standards (PDS4) began, based on lessons learned during twenty years of archiving, current information systems concepts, and the capabilities inherent in the modern internet. PDS4 ensures more consistent standardization, better long-term preservation, and improved automation support \cite{Prockter2021}. Currently, all new datasets submitted to PDS comply with the PDS4 standards, and all existing PDS3 datasets are in the process of being migrated to PDS4. Further details of the history and data standards of the PDS are provided in \citet{Raugh_2021}.

The release of the Memorandum from the President's Office of Science and Technology Policy entitled ``Increasing Access to the Results of Federally Funded Scientific Research'' in 2013 led to the development of the NASA Plan for Increasing Access to the Results of Scientific Research in 2014, the NASA Science Mission Directorate's Strategy for Data Management and Computing for Groundbreaking Science 2019--2024, and the NASA Science Mission Directorate's Policy Document SPD-41: Scientific Information Policy in 2021. Taken together, these documents codify the requirements surrounding the accessibility and preservation of digital data produced by NASA missions, instruments, and projects. The PDS has played a central role in serving as a publicly accessible and appropriate repository for archiving planetary data. The strength of the PDS lies in its long-term stability and the fact that all data are peer reviewed, both to ensure that they conform to the PDS standards and to confirm that they are well-documented and will be usable by future researchers. The PDS is an archive guided by the FAIR principles for scientific data management and stewardship, namely that the data should be findable, accessible, interoperable, and reusable.

After nearly five decades exploring the giant planets of our Solar System, the PDS is now responsible for the curation, preservation, and distribution of a wealth of remote sensing data on the outer planets. This includes data covering a wide range of wavelengths, instrumentation, acquisition modalities, targets, and physical processes. In this paper, we describe examples of such data in %% NJC: Section is fine instead of \S
Section \ref{sec:Data}, discuss the modes through which these data can be accessed in Section \ref{sec:Discoverability}, highlight anticipated future giant planets data and their application to both Solar System and exoplanetary science investigations in Section \ref{sec:Discussion}, and provide conclusions in Section \ref{sec:Conclusion}.

%%%%%%%%%%%%%%%%%%%%%%%%%%%%%%%%%%%%%%%%%%
\section{Giant Planets Data in the PDS} \label{sec:Data}
%Materials and Methods should be described with sufficient details to allow others to replicate and build on published results. Please note that publication of your manuscript implicates that you must make all materials, data, computer code, and protocols associated with the publication available to readers. Please disclose at the submission stage any restrictions on the availability of materials or information. New methods and protocols should be described in detail while well-established methods can be briefly described and appropriately cited.

%Research manuscripts reporting large datasets that are deposited in a publicly available database should specify where the data have been deposited and provide the relevant accession numbers. If the accession numbers have not yet been obtained at the time of submission, please state that they will be provided during review. They must be provided prior to publication.

Remote sensing data relevant to numerous science questions about the giant planets can be found across the PDS, illustrative of both the breadth and depth of the existing giant planets data. With the current PDS4 standards it should be immaterial where the data are actually located since they should be discoverable from any search within the PDS. However, the DNs still play an important and active role in working with and advising new and experienced data providers, data users, and tool developers who interact with data in the PDS.

In the following subsections we provide representative examples of giant planets data that are available from the PDS, organized by science area. {\textbf{This is by no means an exhaustive list.}} %MDPI: Please confirm if the bold is unnecessary and can be removed. The following highlights are the same.
%% NJC: we removed the italics for "representative examples" but we retain the bold in the above sentence as it is a point that we want to strongly emphasize.
 Rather, it is intended to showcase the aforementioned breadth of data and the physical processes they probe. Many instances of giant planet observations in the PDS archive were obtained when the planet was not encountered by the spacecraft, {i.e.}, at distances where the planet is poorly resolved or unresolved by imaging instruments. We highlight here only datasets where the spacecraft was in the vicinity of the giant planet and conducted scientific observations during a flyby, or the planet was the subject of a prolonged orbiter mission. We also provide examples of Earth-based observations of the giant planets that are archived in the PDS.
%with the giant planet as the primary target.
%{\em there are a number of datasets that use the giant planets for distant calibration or distant photometry for limited time... should the scope of the observations be limited to flyby data? If so... this should be stated here as one of the criteria of the showcased data.} Radio Astronomy data are available from Earth. For instance, Voyager started observing Jupiter right after launch. NASA and ESA regularly collect Radio Astronomy data and they are in PDS. Ray}

\subsection{Atmospheric Science}

The most widely recognized attribute of the giant planets is their dynamic atmospheres, including the presence of latitudinally confined cloud bands, long-lasting vortices, discrete convective cloud features, auroral emissions, and evidence of bolide impacts. The PDS archives a broad range of observations that support investigations of these phenomena, including some with temporal coverage spanning several decades that enable studies of seasonal changes in the giant planet atmospheres and comparative studies between the four giant planets of our Solar System.

\subsubsection{Atmospheric Composition}

The chemical composition of gas giant atmospheres can yield critical insights into both their formation and evolution as well as dynamical processes taking place below the cloud tops. Abundance measurements for the most ubiquitous elements (helium, carbon, nitrogen, and sulfur) as compared to the corresponding solar values provide constraints on different planetary formation scenarios \cite{Owen2006}. Global abundances and the spatial distribution of disequilibrium species (e.g., PH$_3$, CO) provide linkages between the atmospheric thermochemistry and the vertical convective motions, which also probes the heat balance of the deep atmosphere.

\begin{figure}[H]
\includegraphics[width=13 cm]{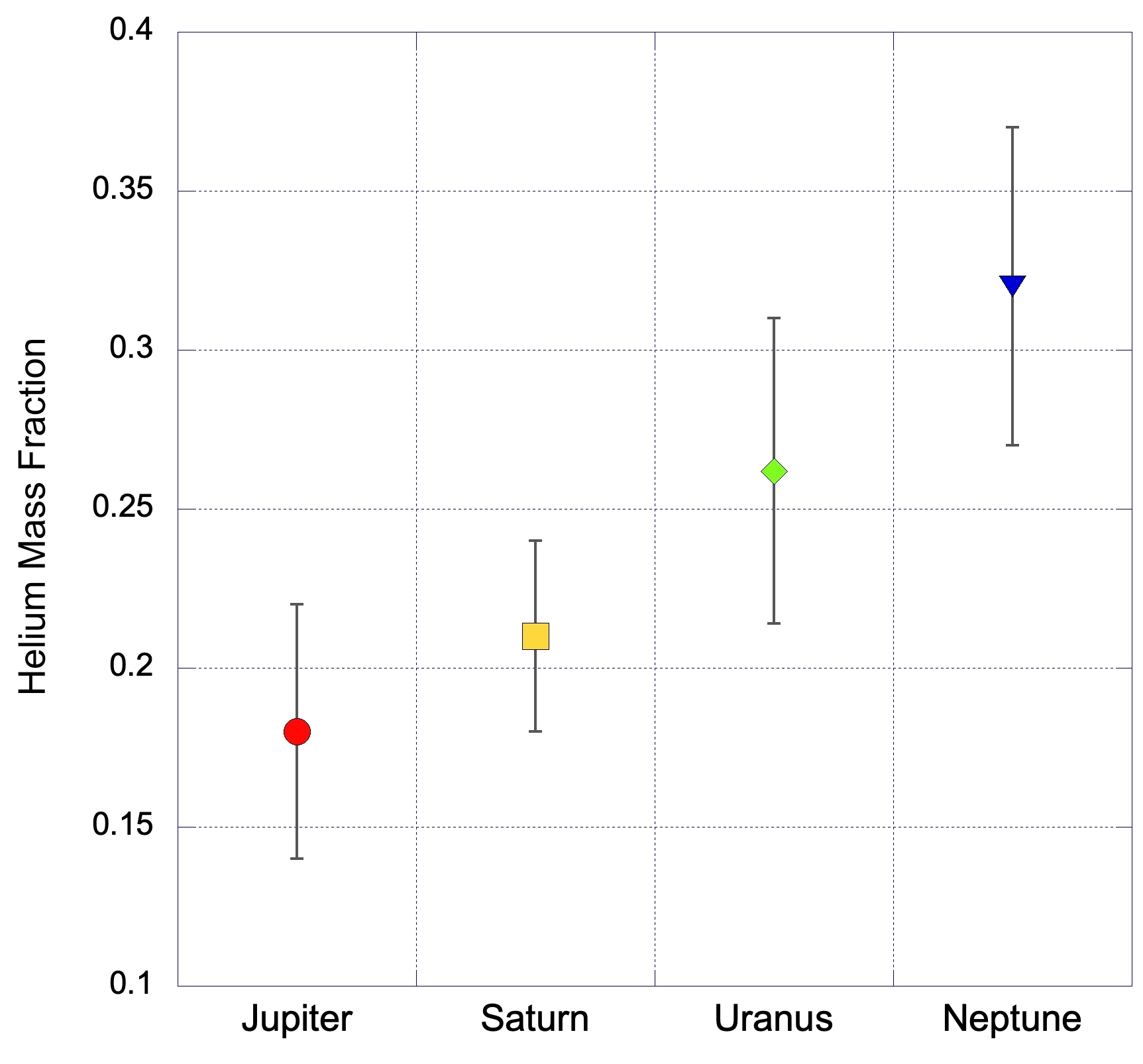}
\caption{Helium abundances of Jupiter \cite{Conrath1984}, Saturn \cite{Conrath2000}, Uranus \cite{Conrath1987}, and Neptune \cite{Conrath1991} derived from Voyager 1 and 2 IRIS observations.}
\label{fig:helium}
\end{figure}

\textls[+10]{Several long-wave infrared (LWIR) spectrometers have flown on NASA missions to the outer planets, including the Infrared Interferometer Spectrometer and Radiometer (IRIS) instrument on Voyagers~1 %MDPI: Please confirm if the italics is unnecessary and can be removed. The following highlights are the same.
%% NJC: italics are unnecessary and we removed them
 and 2 and the Composite Infrared Spectrometer (CIRS) on Cassini. The flyby nature of the Voyager missions resulted in the same instruments being used to probe similar phenomena across all four of the giant planets, which can enable comparative studies. For example, data from the IRIS instruments were used to determine the helium abundance of Jupiter \cite{Gautier1981,Conrath1984}, Saturn \cite{Conrath1984,Conrath2000}, Uranus \cite{Conrath1987}, and \linebreak Neptune \cite{Conrath1991} (Figure \ref{fig:helium}). The 13 year duration of the Cassini mission enabled studies of seasonal variation in Saturn's atmospheric composition and chemistry using the CIRS data \cite{Fouchet2009,Fletcher2010,Fletcher2018}. ({Both the raw and calibrated CIRS data are archived in the PDS and are accessible from}} \linebreak {\url{https://atmos.nmsu.edu/data_and_services/atmospheres_data/Cassini/inst-cirs.html}} \linebreak {and \url{https://pds-rings.seti.org/cassini/cirs/index.html}}, both accessed on 30 October 2022). 

\subsubsection{Global Dynamics}\label{Global}

%The large-scale zonal and meridional wind speeds of the giant planets can be inferred through cloud-tracking techniques. This requires images with appropriate temporal spacing of ideally one to several hours -- a time separation that is too small will not yield measurable motion, whereas a time separation that is too large increases the likelihood that the feature(s) being tracked will evolve and lose their fidelity. Cloud tracking techniques also are optimized when using images of high spatial resolution, which increases the number of features that can be tracked.
%\textcolor{olive}{-GB {\em is not the temporal sampling instrumentally dependant as well? I think the way it is described above is sufficient... otherwise the description can be too specific to the individual case?}} \\
%Every mission to the giant planets was equipped with some form of an imaging camera (Table~\ref{img-tab}), thus there are a wealth of data in the PDS that can be used for studies of giant planet atmospheric dynamics. Studies of long-term ({e.g.} seasonal) variations in giant planet atmospheric dynamics are enabled through analysis of imaging data spanning multiple missions \cite{Simon1999}.

Every mission to the giant planets was equipped with some form of an imaging camera (Table~\ref{img-tab}), resulting in a wealth of data in the PDS that can be used for studies of giant planet atmospheric dynamics. The Voyager flybys revealed the large-scale zonal structure of the atmospheric circulations of Jupiter \cite{Smith1979a,Smith1979b}, Saturn \cite{Smith1981,Smith1982}, Uranus \cite{Smith1986}, and Neptune \cite{Smith1989}, with the Galileo and Cassini orbiters revealing time-dependence and small-scale structure in the wind fields on Jupiter \cite{Vasavada1998c} and Saturn \cite{Porco2005c,Garcia-Melendo2011a} %% NJC: added ", respectively
, respectively. There were also valuable secondary imaging science opportunities during gravity assist flybys as conducted during Jupiter flybys by Cassini \cite{Porco2003a,Li2006a} and New Horizons \cite{Simon2015}.

The large-scale zonal and meridional wind speeds of the giant planets can be inferred through cloud-tracking techniques. This requires images with appropriate temporal spacing of ideally one to several hours: a time separation that is too small will not yield measurable motion, whereas a time separation that is too large increases the likelihood that the feature(s) being tracked will evolve and lose their fidelity. Cloud tracking techniques also are optimized when using images of high spatial resolution, which increases the number of features that can be tracked.

\begin{table}[H] 
\tablesize{\footnotesize}
\caption{Optical %MDPI: We changed link format in table. Please add accessed on date. %% NJC: we added a single "accessed on" statement in the table caption to avoid cluttering up the table
 and near-infrared imagers used for remote sensing of Jupiter (J), Saturn (S), Uranus (U) and Neptune (N), and the Digital Object Identifiers (DOIs) for accessing the data. All links in this table were accessed on 30 October 2022. 
\label{img-tab}}

\begin{adjustwidth}{-\extralength}{0cm}
\centering
\setlength{\cellWidtha}{\fulllength/4-2\tabcolsep-0.3in}
\setlength{\cellWidthb}{\fulllength/4-2\tabcolsep+0.9in}
\setlength{\cellWidthc}{\fulllength/4-2\tabcolsep-1in}
\setlength{\cellWidthd}{\fulllength/4-2\tabcolsep+0.8in}
\scalebox{1}[1] {\begin{tabularx}{\fulllength}
{
>{\PreserveBackslash\raggedright}m{\cellWidtha}
>{\PreserveBackslash\raggedright}m{\cellWidthb}
>{\PreserveBackslash\raggedright}m{\cellWidthc}
>{\PreserveBackslash\raggedright}m{\cellWidthd}}

\toprule

\textbf{Spacecraft}	& \textbf{Instrument} & \textbf{Target}	& \textbf{Access DOIs}\\
\midrule
Pioneer 10, 11 & Imaging Photopolarimeter (IPP) & J, S & not available *\\ \midrule
Voyager 1      & Imaging Science Subsystem (ISS)  & J & \url{https://doi.org/10.17189/1520214}%MDPI: Please add the access date (format: Date Month Year), e.g., accessed on 1 January 2020.
\\
               &                                  & S & \url{https://doi.org/10.17189/1520304}\\ \midrule
Voyager 2      & ISS  & J & \url{https://doi.org/10.17189/1520214}\\
               &      & S & \url{https://doi.org/10.17189/1520304}\\
               &      & U & \url{https://doi.org/10.17189/1520365}\\
               &      & N & \url{https://doi.org/10.17189/1520412}\\ \midrule
Galileo        &  Solid State Imager (SSI) & J & \url{https://doi.org/10.17189/1520425}\\
               & Near Infrared Mapping Spectrometer (NIMS) & J & \url{https://doi.org/10.17189/1520293}\\ \midrule
Cassini        & Imaging Science Subsystem (ISS) & J & \url{https://doi.org/10.17189/1520210}\\
               &                                 & S & \url{https://doi.org/10.17189/1520177}\\ 
               & Visual and Infrared Mapping Spectrometer (VIMS) & J, S & \url{https://doi.org/10.17189/1520275}\\ \midrule
New Horizons   & Long Range Reconnaissance Imager (LORRI) & J & \url{https://doi.org/10.26007/tcne-cm20}\\                
               & Multispectral Visible Imaging Camera (MVIC) & & \url{https://doi.org/10.26007/xmy5-zx84}\\ \midrule
Juno           & JunoCam & J & \url{https://doi.org/10.17189/1520191}\\
\bottomrule
\end{tabularx}}
\end{adjustwidth}

\noindent{\footnotesize{* Pioneer 11 imaging data can be obtained from NSSDC, but not from the PDS.}}
\end{table}

%Multiband imaging observations of Saturn acquired with the Cassini Imaging Science Subsystem (ISS) revealed xxx about Saturn's horizontal wind field (both zonal and meridional... check) \cite{DelGenio2012}.

\subsubsection{Vortices}

\begin{figure}[H]
\includegraphics[width=8 cm]{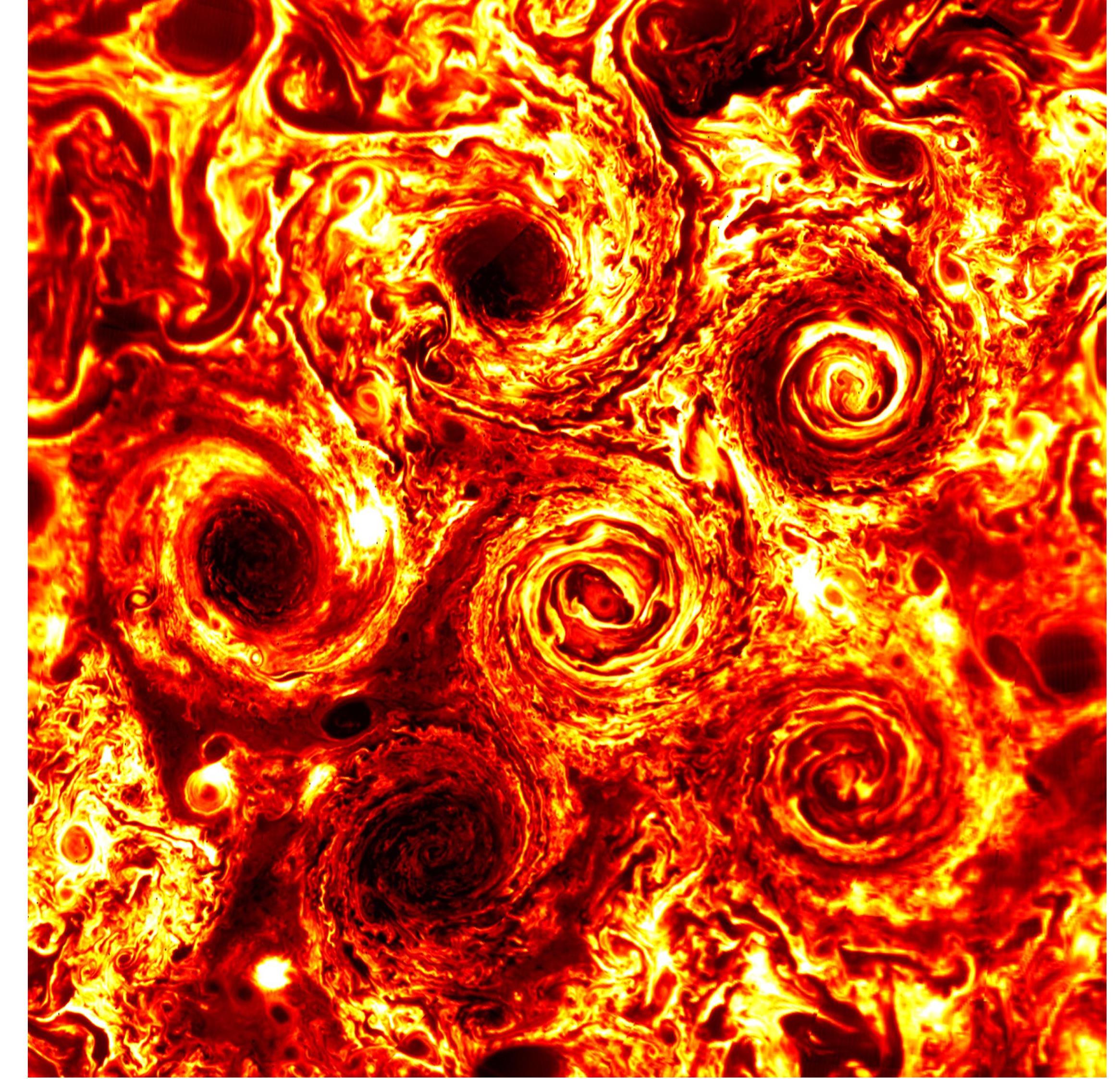}
\caption{Six cyclones in Jupiter's south polar region as revealed by a Juno/JIRAM image taken on February 2, 2017, at a wavelength of roughly 5 microns. Image Credit: NASA/JPL-Caltech/SwRI/ASI/INAF/JIRAM (NASA PIA23556).}
\label{fig:JIRAM}
\end{figure}

In addition to enabling the study of global dynamics of the giant planets, the imaging instruments described in Section \ref{Global} have also been used for detailed studies of discrete vortices. Long-lived storms like Jupiter's Great Red Spot have been characterized in terms of their chemistry and dynamics using observations from several instruments with data archived in the PDS \cite{Simon2014,SanchezLavega2018}. In a more recent NASA mission, Juno's polar orbit has provided unprecedented access to the polar regions of Jupiter, and data from the JIRAM instrument ({\url{https://pds-atmospheres.nmsu.edu/data_and_services/atmospheres_data/JUNO/jiram.html}}, accessed on 30 October 2022), as shown in Figure \ref{fig:JIRAM}, revealed new insights into the formation, evolution, and vertical extent of polar vortex crystals.

\subsubsection{Vertical Profiles}

The variation of atmospheric temperature as well as molecular mole fractions with depth (or pressure) can be inferred by radio or stellar occultations. In the former, the S or X band (or both) radio signals from the spacecraft are the source, and the reduction in signal as it passes from the spacecraft through a giant planet atmosphere to be received on Earth by the Deep Space Network can be analyzed to produce pressure-temperature profiles as well as vertical abundance profiles. For stellar occultations, a star is the source, and the dimming of starlight as measured by remote sensing instruments on board a spacecraft (or on Earth) can yield vertical profiles for atmospheric constituents \cite{Elliot1996}, particularly when components of the occultation data are spectroscopic \cite{Greathouse2010}. The radio science and UV and IR spectroscopic remote sensing instruments on board Voyager 1 and 2, Galileo, Cassini, New Horizons, and Juno have all been used to extract information about the behavior of the giant planet atmospheres with depth. A quick-look view of the types of data archived in the PDS for each giant planet is available through the Integrated Archive pages at the PDS Atmospheres Node. (For example, the Integrated Jupiter Data Archive page (\url{https://pds-atmospheres.nmsu.edu/data_and_services/atmospheres_data/JUPITER/matrices.html}, accessed on 30 October 2022) provides a summary of the remote sensing instruments on board the spacecraft that flew by or orbited Jupiter.)
Additionally, the RMS Node hosts 53 bundles of Earth-based, stellar occultations of the Uranian system ({\url{https://pds-rings.seti.org/pds4/bundles/uranus_occs_earthbased}}, accessed on 30 October 2022). Approximately half of these observations include atmospheric occultations by Uranus. They are presented as time series data, in units of counts-per-time, and are available for users who wish to further analyze them to generate vertical profiles of the Uranian atmosphere.
%At the time of the submission of this paper, they have not been further analyzed to produce vertical profiles of the atmosphere.

Other than occultations, the microwave observations from Juno's Microwave Radiometer (MWR) instrument are sensitive to deeper regions of the Jovian atmosphere (from 0.5 bar to hundreds of bars of pressure) than any previously flown giant planet remote sensing instrument. These data, which are available from the PDS Atmospheres Node ({\url{https://pds-atmospheres.nmsu.edu/data_and_services/atmospheres_data/JUNO/microwave.html}}, accessed on 30 October 2022), are comprised of brightness temperatures in six different radiometric channels. The combination of data from MWR as well as instruments sensitive to Jupiter's upper troposphere ({e.g.}, JIRAM or JunoCam) is a powerful approach for exploring the depths to which atmospheric phenomena such as waves or storms extend \cite{Fletcher2020}.

Atmospheric entry probes are the most direct way to gain information about the vertical structure of the giant planet atmospheres. To date, only the Galileo mission to Jupiter was equipped with a dedicated probe, which entered Jupiter's atmosphere on 7 December 1995. Key science results from the Galileo probe include the findings that Jupiter's zonal winds increased with depth and remained high (and constant) from \linebreak 4 to 21 bars \cite{Atkinson1997},  the relatively cloud-free Jovian ``hot spot'' where the probe entered the atmosphere had an anomalously low water abundance \cite{Niemann1998}, and the probe detected radio signals attributed to the existence of Jovian lightning \cite{Rinnert1998}.
%% NJC: Reworded the following sentence for clarity:
Additionally, at Saturn, when Cassini approached and plunged into the atmosphere, Cassini Radio Science was actively following the spacecraft with multiple Deep Space Network (DSN) receivers until the signal was lost.
These {\em in situ} data, the Galileo Probe and the End-of-Mission raw Cassini radio science data, are archived in the PDS Atmospheres Node ({\url{https://pds-atmospheres.nmsu.edu/data_and_services/atmospheres_data/Galileo/galileo.html} and \url{https://atmos.nmsu.edu/data_and_services/atmospheres_data/Cassini/inst-rss.html}}, both accessed on 30 October 2022). 

\subsubsection{Upper Atmospheres and Aurora}

\begin{figure}[H]
\includegraphics[width=11 cm]{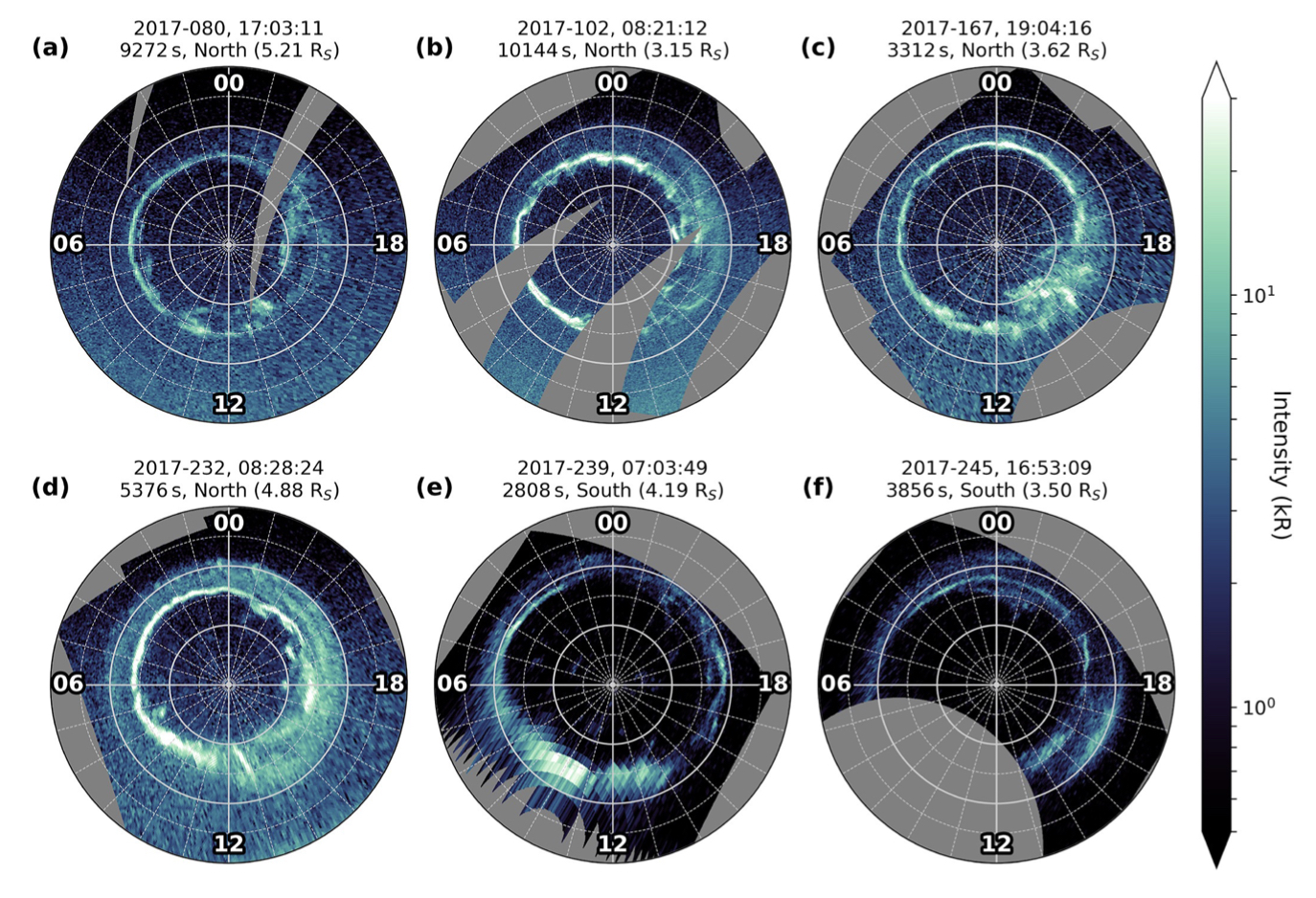}
\caption{UVIS observations of Saturn's northern (\textbf{a}--\textbf{d}) and southern (\textbf{e},\textbf{f}) auroral ovals during the Grand Finale phase of the Cassini mission. Figure reproduced from \citet{Bader2020}.}
\label{aurora-fig}
\end{figure}

Many of the spacecraft destined for the giant planets have been equipped with remote sensing instruments that enable the study of the neutral upper atmospheres, ionospheres, and auroral emissions. The ultraviolet spectrometers listed in Table~\ref{UV-tab} detected Ly$_\alpha$ emission from the giant planets, revealing the complex interactions between each planet's magnetic field and the solar wind and more local sources of charged particles. The Cassini Ultraviolet Imaging Spectrograph (UVIS) provided the most extensive dataset of a giant planet upper atmosphere to date. The UVIS observations of Saturn included solar and stellar occultations in the extreme ultraviolet (EUV) and far ultraviolet (FUV), dayglow spectral images, and images of the magnetosphere that show the vertical profile of neutral hydrogen escaping from Saturn's atmosphere \cite{Nagy2009}. As shown in Figure \ref{aurora-fig}, the Grand Finale phase of the Cassini mission enabled views of Saturn's polar regions of unprecedented detail, revealing a complex structure of Saturn's aurora that relates to the quiet or disturbed nature of the nearby plasma \cite{Bader2020}.

\textls[-15]{In addition to the auroral processes probed by the ultraviolet instruments, the Jovian Infrared Auroral Mapper (JIRAM) instrument on board the Juno spacecraft is sensitive to the 3.3--3.6 $\upmu$m emissions from H$_3^+$, which are used to characterize the morphology of the aurora and the conditions in the magnetosphere that lead to auroral emissions. The JIRAM nadir and limb observations of the H$_3^+$ emissions ({\url{https://pds-atmospheres.nmsu.edu/data_and_services/atmospheres_data/JUNO/jiram.html}}, accessed on 30 October 2022) reveal complex morphology in Jupiter's auroral ovals and an anti-correlation between derived thermospheric temperatures and the H$_3^+$ density, reinforcing that H$_3^+$ emission is an atmospheric cooling mechanism \cite{Dinelli2019}.}

\begin{table}[H] 
\caption{Ultraviolet %MDPI: We changed link format. Please add accessed on date of them. %% NJC: we added a single "accessed on" statement in the table caption to avoid cluttering up the table
 spectrometers used for remote sensing of Jupiter (J), Saturn (S), Uranus (U) and Neptune (N), and the DOIs for accessing the data. All links in this table were accessed on 30 October 2022.
\label{UV-tab}}

\begin{adjustwidth}{-\extralength}{0cm}
\centering
\setlength{\cellWidtha}{\fulllength/4-2\tabcolsep-0.3in}
\setlength{\cellWidthb}{\fulllength/4-2\tabcolsep+0.9in}
\setlength{\cellWidthc}{\fulllength/4-2\tabcolsep-1in}
\setlength{\cellWidthd}{\fulllength/4-2\tabcolsep+0.8in}
\scalebox{1}[1] {\begin{tabularx}{\fulllength}
{
>{\PreserveBackslash\raggedright}m{\cellWidtha}
>{\PreserveBackslash\raggedright}m{\cellWidthb}
>{\PreserveBackslash\raggedright}m{\cellWidthc}
>{\PreserveBackslash\raggedright}m{\cellWidthd}}

\toprule
\textbf{Spacecraft}	& \textbf{Instrument} & \textbf{Target}	& \textbf{Access DOIs}\\
\midrule
Voyager 1  & Ultraviolet Spectrometer (UVS)  & J & \url{https://doi.org/10.17189/4p9r-gc87}\\
           &                           & S & \url{https://doi.org/10.17189/8ads-fr53}\\ \midrule
Voyager 2  & UVS  & J & \url{https://doi.org/10.17189/0w47-dq75}\\
           & & S & \url{https://doi.org/10.17189/emh4-v313}\\
           & & U & \url{https://doi.org/10.17189/fec9-4c64}\\
           & & N & \url{https://doi.org/10.17189/2e17-9r73}\\ \midrule
Galileo &  UVS & J & \url{https://doi.org/10.17189/dv1z-cx79}\\
        & & & \url{https://doi.org/10.17189/8n8q-xf47}\\ \midrule
Cassini & Ultraviolet Imaging Spectrograph (UVIS) & J, S & \url{https://doi.org/10.17189/4be3-xq57}\\
        &          &      & \url{https://doi.org/10.17189/zzgw-f046}\\
        & & & \url{https://doi.org/10.17189/kthj-r777}\\ \midrule
New & Alice Ultraviolet Imaging Spectrograph & J & \url{https://doi.org/10.26007/qfvg-5k41}\\
Horizons  &  Linear Etalon Imaging Spectral Array (LEISA) & J & \url{https://doi.org/10.26007/0cc7-4a49}\\ \midrule
Juno & Ultraviolet Spectrograph (UVS) & J & \url{https://doi.org/10.17189/b29k-pv96}\\
     &  & & \url{https://doi.org/10.17189/c32j-7r56}\\
\bottomrule
\end{tabularx}}
\end{adjustwidth}

\end{table}

\vspace{-12pt} 
%

%\begin{table}[H] 
%\caption{This is a table caption. Tables should be placed in the main text near to the first time they are~cited.\label{tab1}}
%\newcolumntype{C}{>{\centering\arraybackslash}X}
%\begin{tabularx}{\textwidth}{CCC}
%\toprule
%\textbf{Title 1}	& \textbf{Title 2}	& \textbf{Title 3}\\
%\midrule
%Entry 1		& Data			& Data\\
%Entry 2		& Data			& Data\\
%\bottomrule
%\end{tabularx}
%\end{table}
%

\subsubsection{Impacts and Convective Outbursts}

Impact events are ubiquitous throughout the Solar System, but they are challenging to observe without monitoring a target of interest night after night since they are signatures of a stochastic process. However, in the case of Comet Shoemaker-Levy 9 (SL9), whose \linebreak 21 fragments impacted the Jovian atmosphere, observers had more than a year of advanced warning between the time the comet was discovered in March 1993 and the dates on which the fragments were predicted to impact Jupiter (July 1994). Numerous ground-based and space-based assets were used to observe the impacts, deriving information about the chemical composition of both the Jovian atmosphere and the comet fragments, the propagation of waves through the Jovian atmosphere, and the interaction of charged particles with both the magnetosphere and upper atmosphere of Jupiter. The spacecraft data from Galileo and Hubble Space Telescope are archived in the PDS Imaging Node and Atmospheres Node, respectively, and some ground-based observations of the SL9 impacts are also archived in the PDS Atmospheres Node ({\url{https://pds-atmospheres.nmsu.edu/data\_and\_services/atmospheres\_data/catalog.htm\#Jupiter}}, accessed on 30 October 2022).

\begin{figure}[H]
\includegraphics[width=10.2 cm]{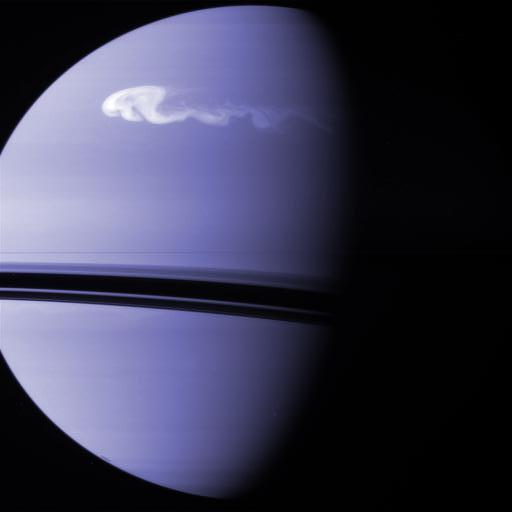}
\caption{\textls[-15]{Cassini ISS image of the 2010 Saturn storm acquired on 24 December 2010. This image was downloaded from the PDS using the RMS OPUS tool (Section 3.1); the tint of this browse image is indicative of the wavelength range of the ISS filter through which the image was acquired (in this case, 429--492 nm).}}
\label{fig:Satstorm}
\end{figure}

\textls[-20]{In addition to impacts, convective outbursts are another episodic phenomenon that can be used to probe vertical transport and moist convection in giant planet atmospheres. Although generally there is little indication {\emph{a priori}} %NJC: this should remain italicized as it is a Latin term
of when such a storm will take place, fortunately these events are often so energetic that their effects are visible for weeks or months after the initial outburst. Such was the case for the Saturn storm of 2010--2011, which was first detected in December 2010 and whose aftermath was detected through August 2011. The Cassini spacecraft was well positioned to acquire observations of the storm with its instrument suite (Figure \ref{fig:Satstorm}), revealing new insights into Saturn's atmospheric structure, wind field, tropospheric cloud composition, stratospheric thermal structure and trace gas composition, ammonia vapor distribution, and associated lightning discharges \cite{SanchezLavega2018}. These data are archived in the PDS Imaging (ISS and VIMS), Atmospheres (CIRS), RMS (ISS, VIMS and CIRS) and PPI (RPWS)~nodes.}

\subsubsection{Ground-Based Photometric Monitoring}

\begin{figure}[H]
\includegraphics[width=13cm]{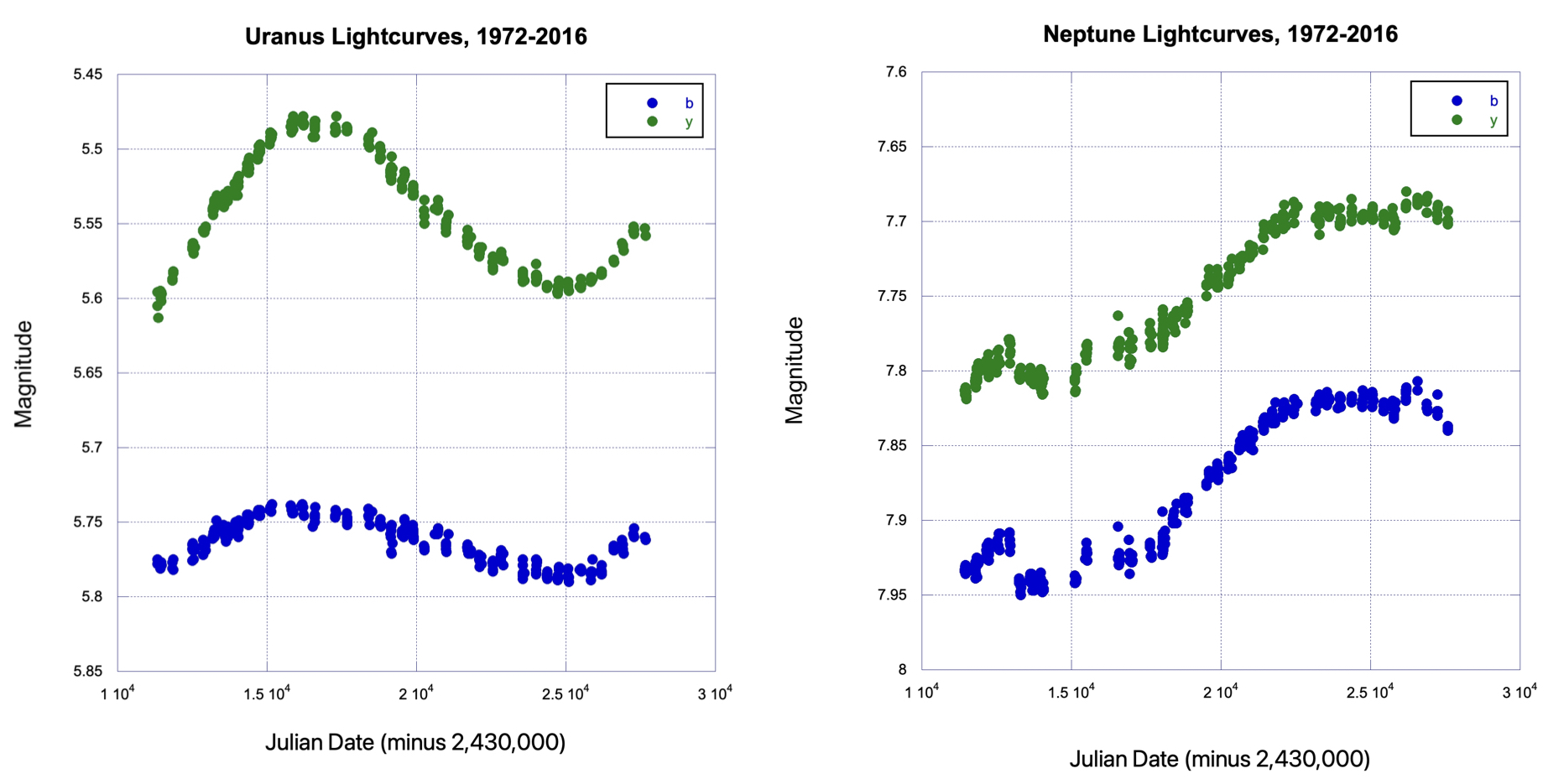}
\caption{Disk %MDPI: please consider to change the hyphen (-) to endash (–) in this figure when it express a kind of range. %% NJC: the software we used to generate these figures does not have the capability to make endash so we left it as is
 integrated light curves of Uranus ({\bf left}) and Neptune ({\bf right}) in the Str\"omgren \textit{b} (blue points) and \textit{y} (green points) filters from 1972 to 2016.}
\label{fig:Lockwoodfig}
\end{figure}

Studies of seasonal changes in the atmospheres of the giant planets are by necessity long-term endeavors, given the orbital periods of the gas giants (approximately 12, 29, 84, and 164 years for Jupiter, Saturn, Uranus, and Neptune, respectively). There are few facilities that have been operating continuously and collecting data in a self-consistent manner that would enable long-term studies of the giant planets; the 21-inch telescope at Lowell Observatory is one such example. Disk-integrated albedos of Uranus and Neptune in the Str\"omgren \textit{b} and \textit{y} filters spanning the period 1972--2016 (Figure \ref{fig:Lockwoodfig}) \cite{Lockwood2019b} are archived in the PDS \cite{Lockwood2019a}. They reveal seasonal changes in the integrated brightness of the ice giant planets as well as short-term brightness changes that are likely due to the appearance and fading of large, discrete atmospheric features.\vspace{-6pt} 

\subsection{Particles and Fields}

Radio frequency emissions from Jupiter are very strong, at times  even stronger than those of the Sun. Radio frequency observations at decametric wavelengths ($\sim$22 MHz) were used to determine that Jupiter had an intrinsic magnetic field \cite{burke1955} before the first \textit{in situ} %MDPI: We removed the italics. Please confirm this revision.
%% NJC: We do not agree with this revision; see comment on first page about this
observations by Pioneer 10. Other Jovian emissions include decametric radiation from synchrotron emissions of trapped electrons in Jovian radiation belts. The first spacecraft to make radio frequency observations were Voyager 1 and 2 in 1979. Both Voyager 1 and Voyager 2 carried a Planetary Radio Astronomy (PRA) experiment and a Plasma Wave Subsystem that measures electric and magnetic field waves in the range of Hz to 10 s of KHz. The radio and plasma wave investigations have probed the strong interaction between the moon Io and the upper atmosphere of Jupiter and have been used to determine the rotation rate of Jupiter. Plasma wave observations have been used to determine the electron density at Jupiter. 

Since Voyager 1 and Voyager 2, the radio or plasma (or both) wave experiments have been flown on four subsequent missions to Jupiter. Table \ref{PPI-tab} gives the spacecraft and instruments, and the Digital Object Identifiers (DOIs) through which the data can be accessed directly from the PDS. The PDS is in the process of migrating all data holdings that were originally archived using the PDS3 metadata standard to the PDS4 standard, which provides a more modern archive structure and richer metadata. In cases where both PDS3 and PDS4 versions of the data are available, we list in Table~\ref{PPI-tab} the access DOIs to the PDS4 data. 

\begin{figure}[H]
\includegraphics[width=13 cm]{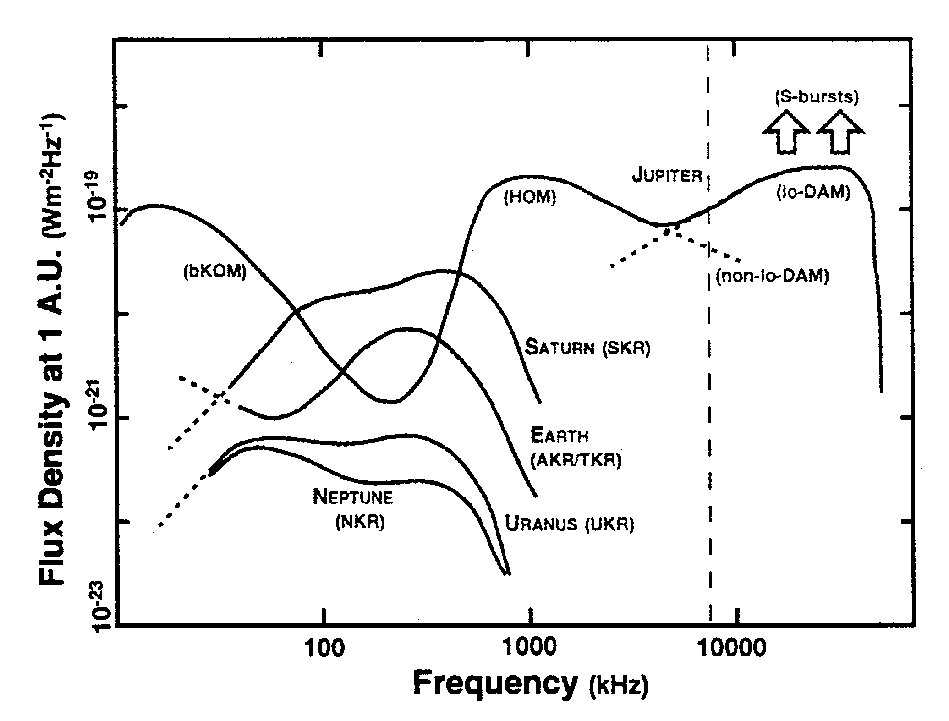}
\caption{The %MDPI: Please use commas to separate thousands for numbers with five or more digits (not four digits) in the picture, e.g., "10000" should be "10,000".
 radio %MDPI: Please change the hyphen (-) into a minus sign ($-$, "U+2212"). e.g., "-1" should be "$-$1".
%% NJC: we cannot change the type or format of the numbers this figure as it is reproduced from another source.
 spectra observed at Jupiter, Saturn, Earth, Uranus, and Neptune \cite{zarka1998}.
\label{radiospectra-fig}}
\end{figure}

Figure~\ref{radiospectra-fig} shows the radio spectra at the giant planets as well as that from Earth. As is the case for Jupiter, electromagnetic wave observations at Saturn have provided insight into the planetary rotation. However, at Saturn the period obtained called planetary period oscillations (PPO) is not the rotation period of the planet. The PPO period drifts over a time scale of years and is different for sources in the north and south (e.g., \cite{Gurnett2007}). Its origin remains an area of intense study. Table~\ref{PPI-tab} lists the radio astronomy and plasma wave data that are available from Saturn.\vspace{-6pt} 

\begin{figure}[H]
\begin{adjustwidth}{-\extralength}{0cm}
\centering
\includegraphics[width=17 cm]{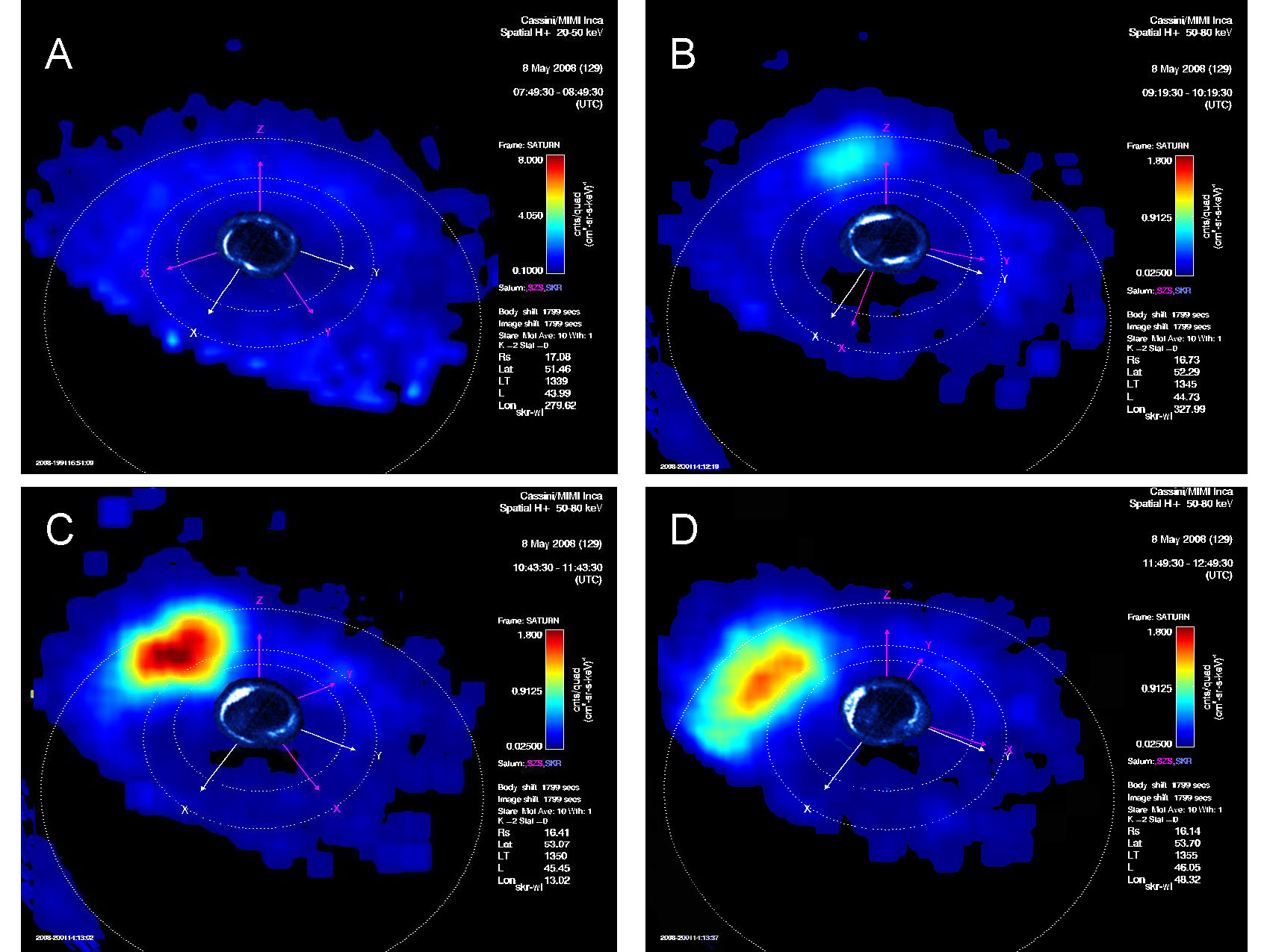}
\end{adjustwidth}
\caption{ENA images of Saturn's magnetospheric ions and auroral EUV emissions during an particle acceleration event on 8 May 2008 ((\textbf{A}) 0749--%MDPI: please consider to change the hyphen (-) to endash (–) in this figure when it express a kind of range. %% NJC: done
0849UT, (\textbf{B}) 0919--1019UT (\textbf{C}) 1043--1143UT, (\textbf{D}) 1141--1149UT). Currents along the magnetic field accelerate the particles producing the aurora. (%MDPI: Another ) missed? Please add it.
%% NJC: added it
Figure courtesy of Donald Mitchell, a movie of the ENA images can be found at \url{https://doi.org/10.17189/1519607}, accessed on 30 October 2022.)}
\label{ENA-fig}
\end{figure}

The Magnetospheric Imaging Instrument (MIMI) on Cassini provided a form of remote sensing of Saturn's magnetosphere not available for the other outer planets. The coexistence of trapped energetic ions and neutral gases in Saturn's magnetosphere allows the process of charge exchange to take place, in which an energetic particle takes an electron from a neutral particle, turning the ion into a neutral particle that is no longer magnetically trapped. The neutral particles can be imaged, providing a picture of the source energetic ion population, in a process called Energetic Neutral Atom (ENA) imaging.  The Ion and Neutral Camera (INCA), was part of Cassini's MIMI package. The INCA camera data are available from the PDS (Table \ref{PPI-tab}), including processed images and movies of the ENA that are included as browse files. In Figure \ref{ENA-fig}, we show a sequence of four ENA images in the magnetosphere along with auroral EUV images in the ionosphere. The images show the injection of ions and the corresponding aurora. Note, the images of the ions map along magnetic field lines to Saturn's aurora and both rotate with Saturn. 

Only Voyager 2 has visited Uranus and Neptune. Both the Planetary Radio Astronomy and the Plasma Wave Subsystem instruments returned data; these are also listed in Table~\ref{PPI-tab}.

Recently, all of the Voyager Plasma Wave Subsystem data have been made available in bundles covering the entire mission by using the PDS4 standard. The newly submitted PDS4 Voyager data are available in the NASA Common Data Format (CDF) format. PDS4 was designed to allow translation from the formats used in the archive to other formats. In space plasma research, the CDF format is widely used in research. PDS developed a version of CDF (called CDF-A) that adheres to PDS4 archival standards. A large number of research tools are available for data archived in CDF-A. The PPI Node is currently translating most of the PDS4 data in the archive into PDS4 documented CDF-A. Both the original PDS4 data and the translated data will be made available to users.  Researchers wanting to use data translated into CDF-A can find the CDF-A standards documents and user guides at \url{https://pds-ppi.igpp.ucla.edu/doc/index.jsp} (accessed on 30 October 2022).

\begin{table}[H]

\caption{Radio %MDPI: we revised the alignment to the unified formats with tables 1 and 2, please note. Can we change all link format to ``https://doi.org/xxxx''? Please add accessed on date of all the links
%% NJC: we added a single "accessed on" statement in the table caption to avoid cluttering up the table. All links were changed to be preceded by https://doi.org/ .
astronomy, plasma wave, and particle data for Jupiter (J), Saturn (S), Uranus (U) and Neptune (N) available in the PDS, and the DOIs for accessing the data. Several of the investigations listed have multiple data sets; in those cases, the range of sequential DOIs is given. All links in this table were accessed on 30 October 2022.
\label{PPI-tab}}

\begin{adjustwidth}{-\extralength}{0cm}
\centering
\setlength{\cellWidtha}{\fulllength/4-2\tabcolsep-0.5in}
\setlength{\cellWidthb}{\fulllength/4-2\tabcolsep+0.7in}
\setlength{\cellWidthc}{\fulllength/4-2\tabcolsep-1in}
\setlength{\cellWidthd}{\fulllength/4-2\tabcolsep+0.8in}
\scalebox{1}[1] {\begin{tabularx}{\fulllength}
{
>{\PreserveBackslash\raggedright}m{\cellWidtha}
>{\PreserveBackslash\raggedright}m{\cellWidthb}
>{\PreserveBackslash\raggedright}m{\cellWidthc}
>{\PreserveBackslash\raggedright}m{\cellWidthd}}

\toprule
\textbf{Spacecraft}	& \textbf{Instrument} & \textbf{Target}	& \textbf{Access DOIs}\\
\midrule
Voyager 1  & Planetary Radio Astronomy Receiver (PRA) %MDPI: Please confirm if the bold is unnecessary and can be removed. The following highlights are the same.
%% NJC: The bold is unnecessary and was removed throughout the table
 & J & \href{https://doi.org/10.17189/1522972}{https://doi.org/10.17189/1522972}\\
       &                            & S & \href{https://doi.org/10.17189/1522966}{https://doi.org/10.17189/1522966}\\
%\midrule
 		   & Plasma Wave Spectrometer (PWS)      & J & \href{https://doi.org/10.17189/1519900}{https://doi.org/10.17189/1519900} -- %MDPI: please consider to change the hyphen (-) to endash (–) in this figure when it express a kind of range. %% NJC: fixed
 		    \href{https://doi.org/10.17189/1519905}{https://doi.org/10.17189/1519905}\\
 		   &                            & S &  \href{https://doi.org/10.17189/1519905}{https://doi.org/10.17189/1519905}\\
                               
   &                            &   & \href{https://doi.org/10.17189/1519927}{https://doi.org/10.17189/1519927} --
                                       \href{https://doi.org/10.17189/1519928}{https://doi.org/10.17189/1519928}\\
                                       &                            &J, S &  \href{https://doi.org/10.17189/wp0z-1c51}{https://doi.org/10.17189/wp0z-1c51} -- \href{https://doi.org/10.17189/g5fy-rz59}{https://doi.org/10.17189/g5fy-rz59}\\
                                       
\midrule
Voyager 2  & PRA	& J & \href{https://doi.org/10.17189/1519955}{https://doi.org/10.17189/1519955}\\
           &                            & S & \href{https://doi.org/10.17189/1520011}{https://doi.org/10.17189/1520011}\\
           &                            & U & \href{https://doi.org/10.17189/1520041}{https://doi.org/10.17189/1520041} --
                                      \href{https://doi.org/10.17189/1520043}{https://doi.org/10.17189/1520043}\\
           &                            & N &  \href{https://doi.org/10.17189/1519984}{https://doi.org/10.17189/1519984} --
                                      \href{https://doi.org/10.17189/1519986}{https://doi.org/10.17189/1519986}\\

%\midrule
 		   & PWS      & J & \href{https://doi.org/10.17189/1519957}{https://doi.org/10.17189/1519957} --
           \href{https://doi.org/10.17189/1519962}{https://doi.org/10.17189/1519962}\\
           &                            & S & \href{https://doi.org/10.17189/1519960}{https://doi.org/10.17189/1519960} --
                                      \href{https://doi.org/10.17189/1519962}{https://doi.org/10.17189/1519962}\\
           &                            & U & \href{https://doi.org/10.17189/1519960}{https://doi.org/10.17189/1519960} --
                                      \href{https://doi.org/10.17189/1519962}{https://doi.org/10.17189/1519962} \\
           &                            &   & \href{https://doi.org/10.17189/1520044}{https://doi.org/10.17189/1520044} --
            \href{https://doi.org/10.17189/1520046}{https://doi.org/10.17189/1520046}\\
           &                            & N & \href{https://doi.org/10.17189/1519987} {https://doi.org/10.17189/1519987}--
            \href{https://doi.org/10.17189/1519989} {https://doi.org/10.17189/1519989}\\
           &                            &J,S,U,N &  \href{	https://doi.org/10.17189/86sw-jn08}{https://doi.org/10.17189/86sw-jn08} -- \href{https://doi.org/10.17189/bwn5-bs17}{https://doi.org/10.17189/bwn5-bs17}\\
\midrule
Ulysses    & Unified Radio and Plasma Wave Experiment (URAP) & J &  \href{https://doi.org/10.17189/1519865}{https://doi.org/10.17189/1519865} --
           \href{https://doi.org/10.17189/1519873}{https://doi.org/10.17189/1519873}\\
\midrule
Galileo    & PWS      & J & \href{https://doi.org/10.17189/1519678}{https://doi.org/10.17189/1519678} --
       \href{https://doi.org/10.17189/1519684}{https://doi.org/10.17189/1519684}\\
\midrule
Cassini    & Radio and Plasma Wave Science (RPWS) & J & \href{https://doi.org/10.17189/1519614} {https://doi.org/10.17189/1519614} --
                \href{https://doi.org/10.17189/1519617}{https://doi.org/10.17189/1519617}\\
           &                               & S & \href{https://doi.org/10.17189/1519610}{https://doi.org/10.17189/1519610} --
                                           \href{https://doi.org/10.17189/1519617}{https://doi.org/10.17189/1519617}\\
           & Magnetospheric Imaging Instrument (MIMI) Imaging Neutral Camera (INCA)      & S & \href{https://doi.org/10.17189/1519607}{https://doi.org/10.17189/1519607}\\
\midrule
New Horizons       & Pluto Energetic Particle Spectrometer Science Investigation (PEPSSI)	&  J &  \href{https://doi.org/10.26007/v0m7-pw47}{https://doi.org/10.26007/v0m7-pw47};
             \href{https://doi.org/10.26007/hggb-f658}{https://doi.org/10.26007/hggb-f658}\\
              & Radio Science Experiment (REX)	&  J &  \href{https://doi.org/10.26007/61tf-4r89}{https://doi.org/10.26007/61tf-4r89};
             \href{https://doi.org/10.26007/f0cb-zn89}{https://doi.org/10.26007/f0cb-zn89}\\
              & Solar Wind Around Pluto (SWAP)	&  J &  \href{https://doi.org/10.26007/dfhh-kn63}{https://doi.org/10.26007/dfhh-kn63};
             \href{https://doi.org/10.26007/q952-d227}{https://doi.org/10.26007/q952-d227}\\
              & Student Dust Counter (SDC)	&  J &  \href{https://doi.org/10.26007/02b6-rm36}{https://doi.org/10.26007/02b6-rm36};
             \href{https://doi.org/10.26007/ke28-ke10}{https://doi.org/10.26007/ke28-ke10}\\
\midrule
Juno       & RPWS	&  J &  \href{https://doi.org/10.17189/1519708}{https://doi.org/10.17189/1519708} --
             \href{https://doi.org/10.17189/1520498}{https://doi.org/10.17189/1520498}\\
%\midrule
%I removed this because the two instruments in PPI PEPSSAI and SWAP are not remove sensing
%New Horizons       & REX, PEPSSI, SWAP, SDC	&  J &  xxx -
%             xxx\\
\midrule
Radio Jove (Earth-based) & Radio Telescope & J & \href{https://doi.org/10.17189/1520498}{https://doi.org/10.17189/1520498} -- 
            \href{https://doi.org/10.17189/1522500}{https://doi.org/10.17189/1522500}\\
\bottomrule
\end{tabularx}}
\end{adjustwidth}
\end{table}

In addition to the spacecraft observations of the giant planets, Jupiter can be readily observed from Earth's surface. In recent years, a worldwide program of continuous monitoring of Jovian radio waves has been maintained. In cooperation with the International Planetary Data Alliance (IPDA), PDS has archived and made available data from the Radio Jove project. The data are available in the archival version of the CDF-A format and in their native format used at radio observatories. The data from Radio Jove are also listed in Table~\ref{PPI-tab}. A unique aspect of the Radio Jove program is that it includes contributions from both professional and amateur observers. An inexpensive receiver that can detect the Jovian emissions is available through the NASA Radio Jove web site ({\url{https://radiojove.gsfc.nasa.gov/}}, accessed on 30 October 2022). 

Planetary particles and fields data are of interest to scientists in both the astronomy and heliophysics communities. In addition to direct access from PPI, the data are being made available through the International Virtual Observatory Alliance's (IVOA) protocol (EPN-TAP) and the Heliophysics API (HAPI). These will enable users in those communities to stream the data using the protocol with which they are familiar. At this writing, all of the Voyager and Cassini data are available through EPN-TAP and the Voyager, Cassini and Galileo data are available through the HAPI server. In addition, PPI supports the TOPCAT (Tool for OPerations on Catalogs %% NJC: replaced "Furthermore," with "And"
And Tables) display system for data using EPN-TAP and %% NJC: removed "the"
Autoplot for data accessed through HAPI.

\subsection{Interiors}

The deep atmospheres and interior structure of the giant planets can be probed through gravity measurements, whereby the returned radio signal from the spacecraft is used to quantify slight deviations in its orbit due to the gravitational influence of the planet on the spacecraft. The Doppler shifts of the signals from the radio science instruments on board Voyager 1 and 2, Cassini, and Juno were measured by the Deep Space Network to probe the gravity fields of the giant planets. For example, recent analyses of measurements made by the Juno Gravity instrument revealed new insights into the depth limits of Jupiter's zonal winds \cite{Kaspi2020} and Jupiter's asymmetric gravity field \cite{Iess2018}. The radio science data for the giant planets are archived in the PDS at the PPI Node (Voyager and Galileo), the Atmospheres Node (Cassini and Juno) and the Small Bodies Node (New Horizons). The Radio Science subnode, housed at the %% NJC: added hyphen between Ring and Moon
Ring-Moon Systems Node, serves all of the PDS and provides expertise on working with radio science data.

Additionally, certain waves in Saturn's rings have been used as records of processes originating in Saturn's interior, similar to the information provided by a terrestrial seismometer (\cite {Hedman22} and references therein). The data facilitating this research is primarily the derived occultation profiles taken from Cassini VIMS ({\url{https://pds-rings.seti.org/viewmaster/volumes/COVIMS_8xxx}}, accessed on 30 October 2022) and RSS ({\url{https://pds-rings.seti.org/viewmaster/volumes/CORSS_8xxx}}, accessed on 30 October 2022). 

%\subsection{Geometrically Calibrated Observations}

%{\em NC: should not this be discussed in 3.2 rather than in Section 2, which is more about the various physical processes being probed with the giant planets remote sensing data?}\\i
%RMS: processed IMG w/enhanced metadata

%{\em MST: We need to do some more thinking about where this goes.  It seems that Section~2 is organized by science discipline and Section~3 is organized by discovery tools.  This topic could be a subsubsection of either, but it should not be a subsection.}

%%%%%%%%%%%%%%%%%%%%%%%%%%%%%%%%%%%%%%%%%%
\section{Data Discoverability and Usability}\label{sec:Discoverability}

%This section may be divided by subheadings. It should provide a concise and precise description of the experimental results, their interpretation as well as the experimental conclusions that can be drawn.

Given the diversity of giant planets data in the PDS, examples of which are described in Section \ref{sec:Data}, accessing and making use of those data might seem like a task that requires advanced knowledge of data processing and visualization, file manipulation, or data engineering. Here we discuss various methods already implemented -- some across the PDS and some at specific Discipline Nodes -- to aid in giant planets data search and discovery.

A new user to the PDS, or an experienced user who is interested in exploring the PDS holdings rather than conducting a focused search for something specific, should start at the PDS homepage ({\url{https://pds.nasa.gov}}, accessed on 30 October 2022). From there, the Data Search function enables users to search by Target ({e.g.}, Uranus System) or Mission ({e.g.}, Voyager). In each case the search returns links to (a) specific search tools -- including those described in the following subsections -- that are most appropriate for finding the data of interest, and (b) resource pages containing more information about the PDS holdings in the area of interest.

Experienced PDS users, or those who know more specifically what kind of giant planets data they are searching for, may wish to begin their search at the relevant DN. There users can make use of sophisticated tools developed by the DNs that enable searching on parameters such as observing geometry, positional information, or science theme. Examples of such tools are described in the following subsections.

\subsection{OPUS and Viewmaster: Search and Browse the RMS Archive}

An enhanced data browser is provided by Viewmaster ({\url{https://pds-rings.seti.org/viewmaster/volumes}}, accessed on 30 October 2022), which allows users to see every data volume in the RMS archive within an intuitively designed framework so that corresponding preview images, diagrams, indices, and documentation are listed in parallel to each data product, and related files can be found straightforwardly. The process of finding and using PDS data archived by the RMS Node is illustrated in Figure \ref{fig:viewmaster_eg} by an example of how a user can generate a spectrum of Saturn's aurora. 

Alongside Viewmaster, seamless cross-mission cross-instrument search for spacecraft and ground-based remote sensing observations is provided by the RMS Node's Outer Planet Unified Search (OPUS) tool ({\url{https://opus.pds-rings.seti.org}}, accessed on 30 October 2022). OPUS uses extensive metadata generated by the RMS Node using the most current SPICE kernels for enhanced search capabilities and provides preview images for an enhanced browsing experience, so users can easily filter down through the available holdings to reach the desired data products. Figure \ref{fig:opus_eg} illustrates how a user could search through OPUS to obtain the same data (and find additional metadata) used to plot the spectrum described in Figure \ref{fig:viewmaster_eg}. OPUS supports full geometric search of the giant planets for Voyager ISS, Cassini CIRS, ISS, UVIS, and VIMS, and New Horizons LORRI.

\vspace{+4pt} 
\begin{figure}[H]
\hspace{-4.5cm} %apologies for this hack
%\hspace{-3.2cm} %apologies for this hack
\centering{
\resizebox{1.2\textwidth}{!}{\input{viewmaster_fig.tex}}} %tikz code
\caption{Step-by-step %MDPI: Please confirm if the accessed on date can be added for links in figure. %% NJC: added accessed on date in caption.
 instructions on how to find and obtain UVIS data from the PDS via RMS Viewmaster for an observation of Saturn's aurora carried out around 03:35 on day 172 of year 2005. All links in this figure were accessed on 30 October 2022.} % Example from UVIS User's Guide, p. 8}
\label{fig:viewmaster_eg}
\end{figure}
\vspace{-4pt} 
\begin{figure}[H]
\hspace{-4.5cm} %apologies for this hack
\centering{
\resizebox{1.2\textwidth}{!}{\input{opus_fig.tex}}} %tikz code	
%\captionsetup{width=\textwidth}
\caption{Step-by-step %MDPI: Please confirm if the accessed on date can be added for links in figure. %% NJC: added accessed on date in caption.
 instructions on how to find and obtain UVIS data from the PDS via RMS OPUS for an observation of Saturn's aurora carried out around 03:35 on day 172 of year 2005. All links in this figure were accessed on 30 October 2022.} 
% Example from UVIS User's Guide, p. 8
\label{fig:opus_eg}
\end{figure}

 %\textit{[[MM: In analogy, M. Aye (2022) describes click-through of MRO example https://www.sciencedirect.com/science/article/pii/B9780128187210000100]]}\\
 
%\textit{What data sets are supported: OPUS supports full geometric search of the giant planets for Voyager ISS, Cassini CIRS, ISS, UVIS, and VIMS, and New Horizons LORRI.}

\subsection{Ephemeris Tools}
RMS also maintains a suite of tools to assist planetary scientists in the planning, acquisition and interpretation of observations ({\url{https://pds-rings.seti.org/tools}}, accessed on 30 October 2022). These tools help users to understand the context of observations, by generating ephemerides for a planet or any of its moons as a function of time, or diagrams of a planetary system at specified times from various viewpoints, or diagrams tracking moons:

\begin{itemize}
\item Planet Viewer generates a diagram showing the appearance of a planetary system at a specified time. Bodies and rings are rendered with terminators and shadows as appropriate. The viewpoint can be Earth’s center, a particular Earth-based observatory, JWST, HST, or a specific planetary spacecraft. Figure~\ref{fig:JupView} shows diagrams from the Jupiter Viewer tool for the same approximate times of Galileo's observations of Jupiter and its four large moons in 1610.
\begin{figure}[H]
\includegraphics[width=5 cm]{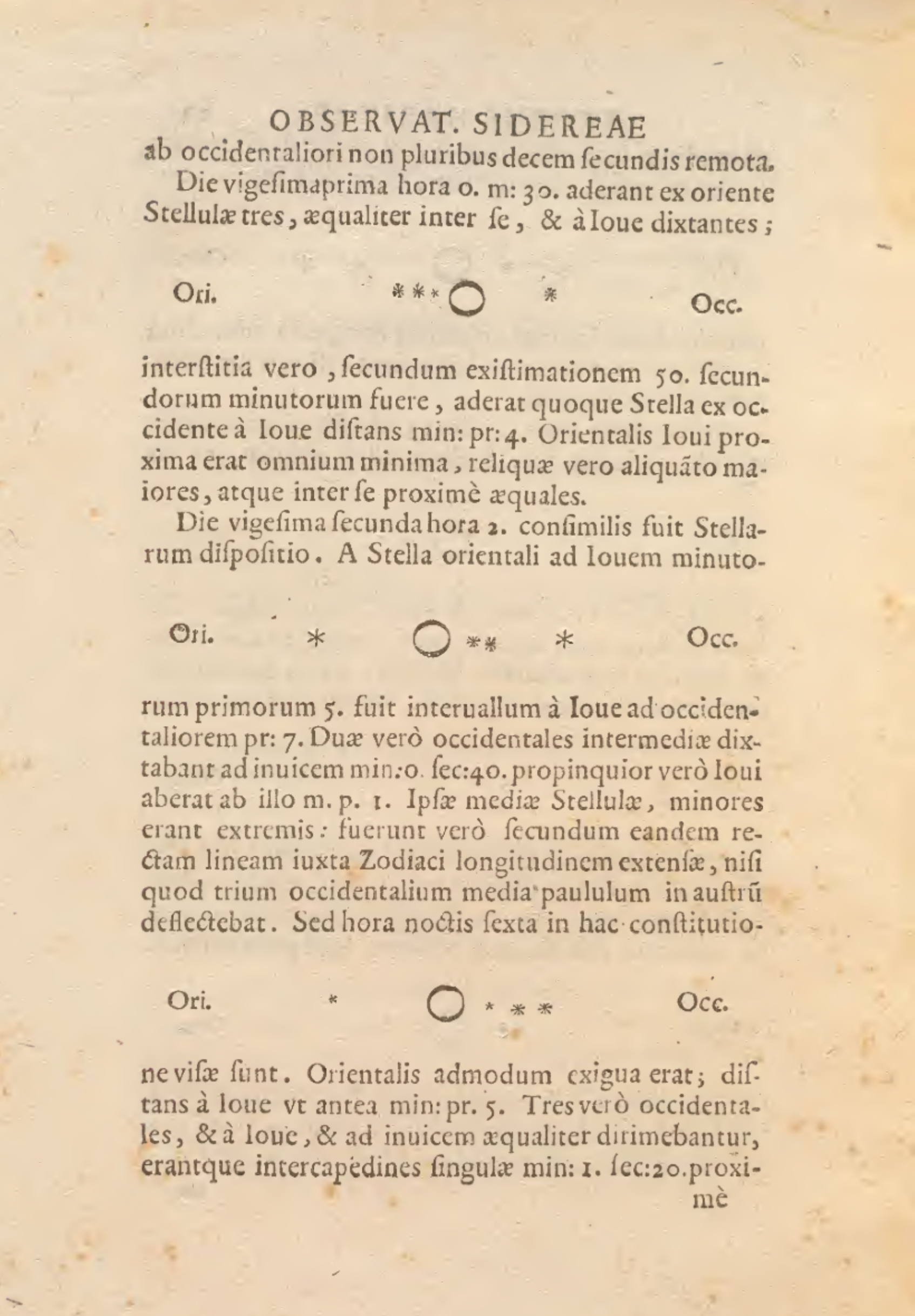}
\includegraphics[width=8.5 cm]{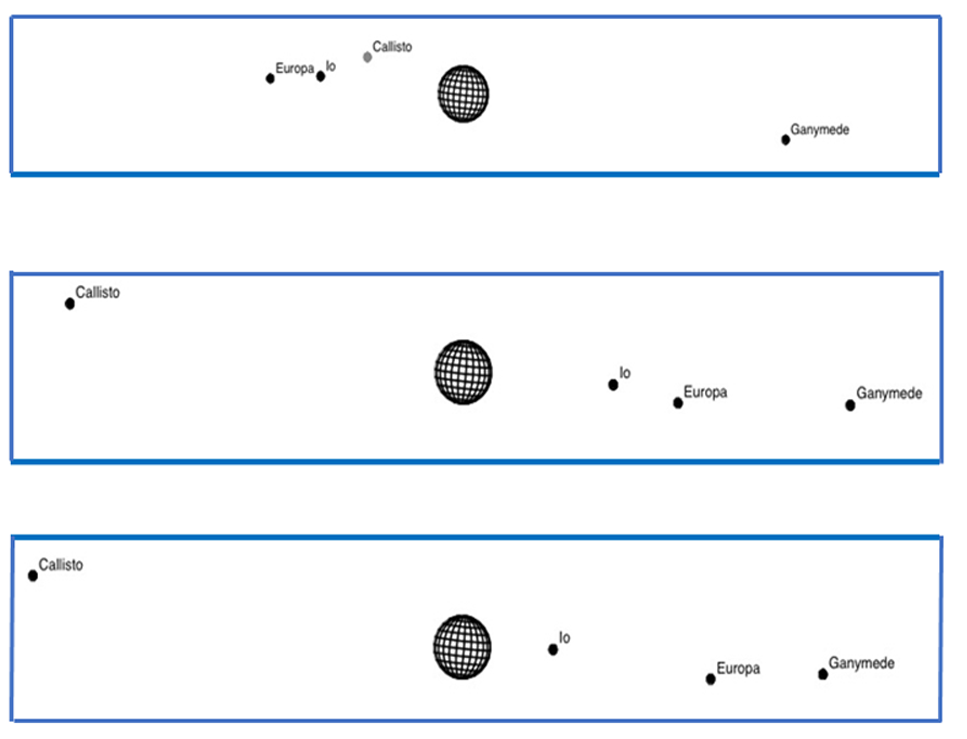}
\caption{Galileo %MDPI: The contents of this figure are not legible. Please replace the image with one of a sufficiently high resolution (min. 1000 pixels width/height, or a resolution of 300 dpi or higher).
%% NJC: we replaced the figure with a higher resolution version
 Galilei's %MDPI: Please remove the non-English term or add a definition for it. %% NJC: this is a proper name, not a non-English term. No change.
 observations \cite{Galilei1610} on the nights of 21 and 22  January 1610 (\textbf{left}) and the corresponding diagrams from the RMS Node's Jupiter Viewer tool (\textbf{right}) for the same approximate observing times.}
\label{fig:JupView}
\end{figure}   

\item Moon Tracker generates a diagram showing the apparent east--west motion of one or more moons relative to the disk of a planet, within a specified time period.
\item Ephemeris Generator generates a table listing of useful information about the viewing geometry for a planet or any of its moons as a function of time. Users are free to specify which of a variety of useful quantities to tabulate (e.g., RA and dec, phase angle, ring opening angle, distance, lunar phase, etc.)
\end{itemize}

The tools are supported for many missions, including Juno, New Horizons, Cassini, Europa Clipper, JUICE, HST, and JWST, and offer temporal coverage between the years 1550 and 2650.
%Figure xxxx shows a comparison of Galileo's observations of Jupiter and the output of the Jupiter Viewer tool for the same (approximate) time.

\subsection{User-Directed Classification via Self-Supervised Learning}

As useful as OPUS and Viewmaster are, the system requires the user to already have knowledge about what data they wish to obtain. The RMS Node is developing a pilot project
%(Smith et al. 2022)
in self-supervised learning \citep{hastie2009elements}, which we hope will eventually result in the ability to add search and classification capabilities based on features in the data.
%(figure?).
For example, users could identify similarities across objects, instruments, or wavelengths to focus on characteristics of interest to their investigations. This is one way in which machine learning methods can speed up the searching process and aid in discovery of relevant data sets.

\subsection{Mission Pages}

The PDS Discipline Nodes host a variety of mission-specific pages to help users access all data from individual missions. Legacy data from missions that were archived using the PDS3 standards are currently being migrated to the new PDS4 archive standard. Experienced users who are familiar with these missions may be better served by accessing the node pages directly; however, new users should start with the mission pages contain a wealth of background and contextual information that will aid them as they begin working with the mission data. Most mission pages serve as portals to instrument-specific data located at the DNs and cross-referenced through these mission pages.

An example set of detailed mission pages are those created for the Cassini mission, which generated 635 GB of science data from 12 instruments while orbiting Saturn for more than 13 years (2004--2017). Near the end of the mission, the Cassini Project in partnership with the PDS developed a set of mission web pages that organized the data by science discipline and included support data and tools to ensure long-term access to the Cassini data by future generations of researchers \cite{Edgington2019}. Mission data archived in the PDS are generally presented to the users by mission, and then organized by instrument. The Cassini web pages maintained by the Atmospheres Node ({\url{https://atmos.nmsu.edu/data\_and\_services/atmospheres\_data/Cassini/sci-saturn.html}}, accessed on 30 October 2022) go beyond this structure to also provide a science-based thematic organization that contains numerous resources for accessing data concerning Saturn's interior and atmosphere. Additional mission pages for Cassini are maintained by the other relevant DNs (RMS ({\url{https://pds-rings.seti.org/cassini}}), accessed on 30 October 2022), SBN, IMG, and PPI).

The Juno mission is an example of an active mission (at the time of this writing), with accumulating datasets that continue to provide a wealth of new observations of the Jovian system. ATM serves as the lead PDS node for the Juno mission and coordinates the archiving of Juno data, which are distributed through the mission web page ({\url{https://pds-atmospheres.nmsu.edu/data\_and\_services/atmospheres\_data/JUNO/juno.html}}, accessed on 30 October 2022). Links to the various instruments take users to corresponding pages at other nodes ({e.g.}, IMG and PPI). These pages are designed to provide information about the mission and instruments to help users find and make use of Juno data.

\subsection{Planetary Image Locator Tool (PILOT)}
The Planetary Image Locator Tool (PILOT) ({\url{http://pilot.wr.usgs.gov}}, accessed on 30 October 2022) is a web-based interface \cite{Bailen:lpsc2013} that provides access to a search tool for several PDS orbital image catalogs indexed by the Unified Planetary Coordinates (UPC) database \cite{Akins:lpsc2009}. The UPC database contains improved geometric, photometric, and positional information about planetary orbital image data, computed using a uniform coordinate system and the most up-to-date SPICE kernels. These improvements result in enhanced image metadata and enable the user's ability to identify desired images, extending beyond those provided by the PDS. Precise searches can be performed within PILOT on individual metadata fields stored in the UPC database. In addition, PILOT has access to browse images, links to the raw images, and label information that can help scientists and cartographers perform further investigations on the images. Although all data in the UPC and PILOT must be supported by a camera model that describes detailed instrument and target geometric behavior during image acquisition, together these services address $\sim$90\% of the orbital image data for which Imaging node is responsible to the PDS. PILOT not only allows users to download select images either individually or through provided scripts to pull large sets of images, it also provides direct access to the Map Projection on the Web (POW) online tool. PILOT was developed and is maintained by the Cartography and Imaging Sciences Node to complement other node delivery services.

\subsection{Map Projection on the Web (POW)}

The Map Projection on the Web (POW) ({\url{https://astrocloud.wr.usgs.gov/}}, accessed on 30 October 2022) service \cite{Hare:lpsc2013,Akins:planetdata2015} allows users to convert raw PDS images to science-ready, map-projected products. POW integrates PILOT \cite{Bailen:lpsc2013} and the UPC \cite{Akins:lpsc2009}, the Integrated Software for Imagers and Spectrometers (ISIS) package ({\url{https://isis.astrogeology.usgs.gov/}}, accessed on 30 October 2022) \cite{Sides:lpsc2017}, Geospatial Data Abstraction Library (GDAL) ({\url{https://gdal.org/}}, accessed on 30 October 2022) \cite{Hare:lpsc2008}, and the Astrogeology cluster for its processing and delivery needs. POW provides users with calibrated cartographic images that can be used readily for analysis of atmospheric dynamics, cloud morphology and change detection, merging of dissimilar instrument images, analysis in a Geographic Image System (GIS) and use in a host of other scientific applications (e.g., ArcMAP, ENVI, Matlab, JMARS, QGIS, etc.). POW allows researchers to make use of a wealth of PDS science data without having to install or learn how to run ISIS and to benefit from a recommended processing pipeline as defined by USGS and the instrument teams. This service can also be used as a learning tool or an introduction to ISIS for those who would like to run it locally because the ISIS commands will be logged and delivered to the user. Using the POW front-end, a user is allowed to (1) select and submit a list of up to 250 PDS raw images, (2) define an output map projection and its parameters (e.g., Polar Stereographic, Sinusoidal), (3) define the output bit type (8, 16, or 32 bit), and (4) select an ISIS or PDS output format or a more standardized geospatial format such as GeoTiff, PNG, or JPEG. Conversion to the various supported image formats will be completed using the GDAL, which passes all cartographic information into the output format.

\subsection{Planetary Image Atlas}
The Atlas ({\url{https://pds-imaging.jpl.nasa.gov/Atlas}}, accessed on 30 October 2022) \cite{Padams:planetdata2017} provides access to the full collection of PDS-IMG data available in online holdings and data node catalogs. The Atlas uses faceted navigation to support searches on common search criteria such as mission name, instrument name, target, product type, observation/illumination geometry metadata, geographic coordinates, time constraints, etc. The Atlas also allows users to search images based on their ``content'' (i.e., moons, rings, clouds) through the use of deep convolutional neural networks to classify images \cite{Altinok:lpsc2014,Stanboli:planetdata2017}.  

\section{Discussion}\label {sec:Discussion}

%Authors should discuss the results and how they can be interpreted from the perspective of previous studies and of the working hypotheses. The findings and their implications should be discussed in the broadest context possible. Future research directions may also be highlighted.

Looking forward, the PDS will be the primary repository for giant planets data from several upcoming missions and derived datasets, as well as supporting research conducted to aid in the interpretation of the remotely sensed giant planets data already archived in the PDS.

\subsection{Future Mission Data}
Juno began its extended mission phase in July 2021. The prime mission data, along with the accumulating data from the extended phase, will continue to be available from the PDS ATM, RMS, PPI, and IMG nodes. Higher order data products resulting from the end of the prime mission such as gravity model harmonic coefficients, auroral UV maps, auroral H$_{3}^{+}$ maps, magnetic field models among other products will also be available from the aforementioned nodes as they complete the peer review process and become ingested into the archive.

Beyond Juno, NASA's Europa Clipper and ESA's JUpiter ICy moons Explorer (JUICE) missions will conduct detailed investigations of several of Jupiter's icy satellites. If observations of Jupiter are also made, those data will be archived in the PDS in the relevant node(s) (or linked to from the PDS in the case of the JUICE data, whose primary archive will be ESA's Planetary Science Archive). 

\subsection{Data Derived from Analysis of Remote Sensing Observations}
% prob do not need to include all of these; can synthesize into a few areas; need to provide citations!
The scientific impact of giant planets remote sensing data in the PDS is felt long after the end of a mission. In some cases the data themselves are used for investigations long after their acquisition, making use of new analysis techniques, tools, or interpretations. In other cases the data are used to inspire new, related projects that enhance the interpretation of the original data. In these cases, the resulting data are often referred to as {\em derived data}, and ideally they should %% NJC: removed extra "be"
also be archived in the PDS to optimize their discoverability and usability.

\textls[-15]{There are numerous examples of derived data relevant to the giant planets that are archived in the PDS, for example, Saturn zonal wind profiles derived from Cassini ISS observations \cite{GarciaMelendo2011} and vertical profiles of Saturn's thermospheric temperature and H$_2$ densities derived from the analysis of Cassini UVIS and CIRS stellar occultation observations \cite{Koskinen2018}. With the greater demand for archiving derived data products now being placed on researchers funded by NASA's Research and Analysis programs, the PDS expects to receive and archive numerous additional derived data products relevant to the giant planets. Both future data users as well as data proposers are encouraged to turn to the PDS as a logical repository for their derived data archiving needs.}

%In Table~\ref{derived-tab}, we list just a small sample, to indicate the breadth and scope of the derived data in the PDS.

%\begin{table}[H] 
%\caption{Examples of derived giant planets data in the PDS. \label{derived-tab}}
%\newcolumntype{C}{>{\centering\arraybackslash}X}
%\begin{tabularx}{\textwidth}{llll}
%\toprule
%\textbf{Target}	& \textbf{Description}	& \textbf{In support of} & %\textbf{Citation}\\
%\midrule
%J, S, U, N	& cruise data	& Voyager & N/A\\
%J		& ground-based spectra and IR imaging & Galileo, Juno		& N/A\\
%J & temporal variability in magnetosphere & Juno & N/A\\
%S & T, $\rho$ thermosphere profiles & Cassini & xxx\\
%\bottomrule
%\end{tabularx}
%\end{table}
%

%Voyager - new PDART data? \textit{[[Is this the RAV1CIUN project?  Or something else?  If it is, MST can add a sentence]]}\\
%Jupiter ground-based, long-baseline support observations\\ 
%oscillation and atmospheric dynamics ground-based observations of the gas giants\\
%Juno observations of temporal variability in Jupiter’s magnetosphere\\
%Multiple mission synthesis for atmospheric dynamics of Jupiter\\
%Magnetic effects on plasma densities in Jupiter’s ionosphere\\
%Juno-derived dynamic tides for Jupiter\\
%Cloud structure and composition variations with discrete features and storms on Saturn\\
%Cassini-derived dynamics of Saturn’s relatively deep atmosphere\\
%Cassini observations of neutral gas in the inner Saturn magnetosphere\\
%Small-scale atmospheric features of Saturn (Slepian functions to interpret Cassini gravity data)\\
%Cassini CIRS interpreted with photochemical modeling\\

\subsection{Data Generated in Support of Giant Planet Studies}

Laboratory experiments involving gases at or near conditions appropriate for the giant planets are required for the interpretation of remote sensing observations of the giant planets. For example, interest in hydrocarbons in the upper atmospheres of Jupiter and Saturn (and Titan) drove the need for better chemistry data to understand the behavior of these species in such extreme conditions \cite{Bernath2022}. The PDS archives laboratory data alongside the relevant mission and ground-based observatory data to aid in the advancement of the science of atmospheric chemistry in the outer Solar System.

Computational models and simulations are another critical tool used for the interpretation of remotely sensed giant planets data. Although model outputs are generally not archived in the PDS as they do not conform to PDS standards, several PDS nodes have implemented -- or are now implementing -- annex-style repositories to meet the data accessibility of the planetary modeling communities \cite{Newman2021}. For example, the PPI Node maintains an annex ({\url{https://pds-ppi.igpp.ucla.edu/search/annex/}}, accessed on 30 October 2022) for models describing the Jovian current sheet and the global magnetic field of Saturn. The Atmospheres Node is in the process of implementing an atmospheric modeling annex that will be used to preserve outputs from atmospheric dynamical models, including those relevant to the giant planets.

\subsection{Connections to Other Archives}

In 2020 NASA chartered the Planetary Data Ecosystem (PDE) Independent Review Board (IRB) to provide a holistic evaluation of the planetary science community's access to and use of planetary data. The PDE IRB report provided scores of findings and recommendations aimed at fostering the development of the ecosystem and addressing barriers to data preservation and the use of planetary data \cite{PDEIRB2021}. The recognition that there is an entire ecosystem surrounding access to and use of planetary data was a critical outcome of the PDE IRB effort, as well as the need to improve the connectivity of the different elements of the ecosystem.

For example, data from the Hubble Space Telescope (HST) have led to many significant discoveries related to the giant planets during its 32-year (and counting) mission.  These include 
%spectacular aurorae in Jupiter's atmosphere [Nichols et al. 2016]
the appearance and disappearance of dark vortices in the atmospheres of Uranus and Neptune \cite{Hsu2019},
the unique nature of Saturn's aurorae \cite{Clarke2005},
%unexpected changes in Neptune's Great Dark Spot and the birth of its ``Dark Spot Jr.'' [Wong et al. 2021], 
the long-term studies of the shrinking of Jupiter's Great Red Spot \cite{Simon2014}, and the changes in Jupiter's zonal wind profile \cite{Tollefson2017}. The HST data, which are archived at the Mikulski Archive for Space Telescopes (MAST, {\url{https://archive.stsci.edu/hlsp/search.php}}, accessed on 30 October 2022), contains thousands of observations of the giant planets, but they are not integrated into PDS and are not easily searchable by planetary scientists. %% NJC: removed the rest of this paragraph: Solar System \hl{targets}. ({Although MAST does contain metadata recording Solar System objects that were an observer's intended target, its search based on sky coordinates cannot (for example) capture Solar System objects that may serendipitously appear in HST's field of view}).

The RMS Node is nearing completion of a pipeline to fully incorporate HST data identified as relevant to the Solar System into PDS4, with fully searchable OPUS metadata. The RMS pipeline can automatically query MAST, identify Solar System observations by a variety of attributes, retrieve the relevant files, and construct a corresponding PDS4 bundle that includes planetary metadata. % We are also developing an HST dictionary in the PDS4 IM, which includes definitions for numerous terms derived from the MAST archive files. 
When this project is complete, OPUS users will be able to access HST data based on a full set of Solar System-focused search attributes.  In the meantime, RMS currently maintains a large collection of “placeholder” HST volumes (in PDS3 format). These contain browse products and metadata, but they redirect users to MAST to retrieve the actual data files. 
%We are hopeful that
The foundation laid by this project will also enable the PDS to give similar treatment to JWST data. 

In an effort to further improve interoperability, the PDS has played a major role in the development of the International Planetary Data Alliance (IPDA, {\url{https://planetarydata.org}}, accessed on 30 October 2022), whose goal is to facilitate access to and the exchange of planetary science data that are managed by a number of international space agencies. Enabling broad, international access to planetary data requires the use of common standards across the archives of different agencies (PDS4 has been adopted as the current standard for IPDA) and clear linkages among the data archives. Although the giant planets have yet to be explored by spacecraft managed by agencies other than NASA, that will change with ESA's JUICE mission, and the PDS will play a key role in ensuring transparent access to the JUICE data from the PDS.

Finally, with the discovery of thousands of giant planets in exoplanetary systems, there has been a tremendous increase in the number of studies that treat the giant planets of our Solar System as analogs for advancing our understanding of exoplanets. Current and future data in the PDS are vital to such studies, including for example full disk albedos of the giant planets \cite{Karkoschka1994,Karkoschka1998}, full-disk images of the giant planets acquired at a wide range of phase angles \cite{Mayorga2016}, and laboratory measurements made at giant planet and exoplanetary analog conditions \cite{Bernath2022}. Efforts are currently underway within NASA's Science Mission Directorate to enhance interoperability across the archives of the Astrophysics, Heliophysics, Planetary Science, and Earth Science Divisions. Exoplanet studies, which are heavily interdisciplinary in nature, are a motivating driver for such efforts and potential use cases are currently being explored.

%%%%%%%%%%%%%%%%%%%%%%%%%%%%%%%%%%%%%%%%%%
\section{Conclusions}\label {sec:Conclusion}

The giant planets data within the PDS represent a record of an incredible human achievement: spacecraft exploration of the outer Solar System aimed at answering fundamental science questions concerning the origin and evolution of planets. The data also serve as the scientific legacy of numerous researchers who have contributed data they collected, analyzed, and used to generate new knowledge about the workings of \linebreak the giant planets.

The data in the PDS are preserved using state-of-the-art information technology, with tools available to aid in their discoverability and usability. The fact that data from NASA's earliest days of giant planet exploration are still being used today to obtain new insights about the workings of the gas giants \cite{Gupta2022} demonstrates their long-lasting utility and benefit to science. The PDS will continue to uphold its role as NASA's preeminent archive for giant planets data while evolving to meet modern data archiving challenges and the changing needs of its user base.

%%%%%%%%%%%%%%%%%%%%%%%%%%%%%%%%%%%%%%%%%%
%\section{Patents}

%This section is not mandatory, but may be added if there are patents resulting from the work reported in this manuscript.

%%%%%%%%%%%%%%%%%%%%%%%%%%%%%%%%%%%%%%%%%%
\vspace{6pt} 

%%%%%%%%%%%%%%%%%%%%%%%%%%%%%%%%%%%%%%%%%%
%% optional
%\supplementary{The following are available online at \linksupplementary{s1}, Table S1: Radio astronomy, plasma wave, and particle data for the giant planets available in the PDS.}


%%%%%%%%%%%%%%%%%%%%%%%%%%%%%%%%%%%%%%%%%%
\authorcontributions{Conceptualization, N.J.C., M.S.T., R.J.W. and M.K.G.; writing---original draft preparation, N.J.C., M.S.T., M.J.T.M., R.J.W., L.F.H., L.D.V.N., J.J.B., J.M.B.; writing---review and editing, M.K.G.; visualization, N.J.C., M.J.T.M., R.J.W., M.K.G. All authors have read and agreed to the published version of the manuscript.}

%\authorcontributions{Conceptualization, NJC, MST, RJW, and MKG; writing---original draft preparation, NJC, MST, MJTM, RJW, LFH, LDVN, JJB, JMB; writing---review and editing, MKG; visualization, NJC, MJTM, RJW, MKG. All authors have read and agreed to the published version of the manuscript.}

\funding{This %MDPI:Please add: ``This research received no external funding'' or ``This research was funded by NAME OF FUNDER grant number XXX.'' and  and ``The APC was funded by XXX''. Check carefully that the details given are accurate and use the standard spelling of funding agency names at \url{https://search.crossref.org/funding}, any errors may affect your future funding.
%% NJC: This is already done; no change.
 research was funded by the Planetary Science Division of NASA's Science Mission Directorate through award numbers 80NSSC22M0020 (ATM), 80NSSC22M0022 (RMS), 80NSSC22M0023 (PPI), 80HQTR22TA003 (IMG), and 80NSSC22M0024 (SBN).}

%\institutionalreview{In this section, please add the Institutional Review Board Statement and approval number for studies involving humans or animals. Please note that the Editorial Office might ask you for further information. Please add ``The study was conducted according to the guidelines of the Declaration of Helsinki, and approved by the Institutional Review Board (or Ethics Committee) of NAME OF INSTITUTE (protocol code XXX and date of approval).'' OR ``Ethical review and approval were waived for this study, due to REASON (please provide a detailed justification).'' OR ``Not applicable'' for studies not involving humans or animals. You might also choose to exclude this statement if the study did not involve humans or animals.}

%\informedconsent{Any research article describing a study involving humans should contain this statement. Please add ``Informed consent was obtained from all subjects involved in the study.'' OR ``Patient consent was waived due to REASON (please provide a detailed justification).'' OR ``Not applicable'' for studies not involving humans. You might also choose to exclude this statement if the study did not involve humans.

%Written informed consent for publication must be obtained from participating patients who can be identified (including by the patients themselves). Please state ``Written informed consent has been obtained from the patient(s) to publish this paper'' if applicable.}

\dataavailability{All of the data described in this paper are publicly available through the NASA Planetary Data System ({\url{https://pds.nasa.gov}}, accessed on 30 October 2022).

%In this section, please provide details regarding where data supporting reported results can be found, including links to publicly archived datasets analyzed or generated during the study. Please refer to suggested Data Availability Statements in section ``MDPI Research Data Policies'' at \url{https://www.mdpi.com/ethics}. You might choose to exclude this statement if the study did not report any data.} 

\acknowledgments{Any %MDPI:In this section, you can acknowledge any support given which is not covered by the author contribution or funding sections. This may include administrative and technical support, or donations in kind (e.g., materials used for experiments). %% NJC: no change
use of trade, firm, or product names is for descriptive purposes only and does not imply endorsement by the U.S. Government. The authors thank R. Beebe for a pre-submission review of the manuscript.}

\conflictsofinterest{The authors declare no conflict of interest. The funders had no role in the design of the study; in the collection, analyses, or interpretation of data; in the writing of the manuscript, or in the decision to publish the results.}

\newpage
%% Optional
%\sampleavailability{Samples of the compounds ... are available from the authors.}

%%%%%%%%%%%%%%%%%%%%%%%%%%%%%%%%%%%%%%%%%%
%% Only for journal Encyclopedia
%\entrylink{The Link to this entry published on the encyclopedia platform.}

%%%%%%%%%%%%%%%%%%%%%%%%%%%%%%%%%%%%%%%%%%
%% Optional
\abbreviations{Abbreviations}{
The following abbreviations are used in this manuscript:\\

\noindent 
\begin{tabular}{@{}ll}
API & Application Programming Interface\\
ATM & Atmospheres Node\\
CDF & Common Data Format\\
CDF-A & Common Data Format Archival\\
CIRS & Composite Infrared Spectrometer (Cassini)\\
CODMAC & Committee on Data Management and Computation\\
Dec & Declination\\
DN & Discipline Node\\
DOI & Digital Object Identifier\\
%% NJC: added one more:
DSN & Deep Space Network\\
%EDR & Experimental Data Record\\
ENA & Energetic Neutral Atom\\
EPN-TAP & EuroPlaNet Table Access Protocol\\
ESA & European Space Agency\\
EUV & Extreme Ultraviolet spectrographic channel of the UVIS instrument (Cassini)\\
FAIR & Findability, Accessibility, Interoperability, and Reuse of digital assets\\
FUV & Far Ultraviolet spectrographic channel of the UVIS instrument (Cassini)\\
GDAL & Geospatial Data Abstraction Library\\
GeoTiff & a standard geospatial format\\
GIS & Geographic Information System\\
HAPI & Heliophysics Application Programming Interface\\
HST & Hubble Space Telescope\\
IMG & Cartography and Imaging Sciences Node\\
INCA & Ion and Neutral Camera (Cassini)\\
IPDA & International Planetary Data Alliance\\
IPP & Imaging Photopolarimeter (Pioneer 10, 11)\\
IR & Infrared\\
IRB & Independent Review Board\\
IRIS & Infrared Interferometer Spectrometer and Radiometer (Voyager 1,2)\\
ISS & Imaging Science Subsystem (Cassini, Voyager 1, 2)\\
IVOA & International Virtual Observatory Alliance\\
JIRAM & Jovian Infrared Auroral Mapper (Juno)\\
JPEG & a common image format developed by the Joint Photographic Experts Group\\
JUICE & JUpiter ICy moons Explorer (European Space Agency mission)\\
JWST & James Webb Space Telescope\\
ISIS & Integrated Software for Imagers and Spectrometers\\
LORRI & Long Range Reconnaissance Imager (New Horizons)\\
LWIR & Long-Wave Infrared Spectrometer (several missions)\\
MAST & Mikulski Archive for Space Telescopes\\
MIMI & Magnetospheric Imaging Instrument (Cassini)\\
MVIC & Multispectral Visible Imaging Camera (New Horizons)\\
MWR & MicroWave Radiometer (Juno)\\
NASA & National Aeronautics and Space Administration\\
NIMS & Near Infrared Mapping Spectrometer\\
NSSDC & National Space Science Data Center\\
OPUS & Outer Planet Unified Search\\
PDE & Planetary Data Ecosystem\\
PDS & Planetary Data System\\
PDS3 & Governing standards of the PDS from the mid-1990s until the adoption of PDS4.\\
PDS4 & Current governing standards of the PDS, adopted in 2005\\
PI & Principal Investigator\\
PILOT & Planetary Image Locator Tool\\
PNG & Portable Network Graphic, a common image format\\
POW & Map Projection on the Web\\
PPI & Planetary Plasma Interactions Node
\end{tabular}}

\noindent 
\begin{tabular}{@{}ll}

PPO & Planetary Period Oscillations\\
PRA & Planetary Radio Astronomy (Voyager 1, 2)\\
PWS & Plasma Wave Subsystem (Voyager 1, 2)\\
RA & Right Ascension\\
RMS & Ring-Moon Systems Node\\
RSS & Radio Science Subsystem (Cassini)\\
RSSN & Radio Science Sub-Node\\
RPWS & Radio and Plasma Wave Science (Cassini)\\
SBN & Small Bodies Node\\
SL9 & Comet Shoemaker-Levy 9\\
SPICE & Observation geometry information system for planetary spacecraft\\
SSI & Solid State Imager (Galileo)\\
TOPCAT & Tool for OPerations on Catalogs %% NJC: replaced "Furthermore," with "And"
And Tables\\
UPC & Unified Planetary Coordinates\\
URAP & Unified Radio and Plasma Experiment\\
USGS & United States Geological Survey\\
UV & Ultraviolet\\
UVIS & Ultraviolet Imaging Spectrograph (Cassini)\\
VIMS & Visual and Infrared Mapping Spectrometer (Cassini)
\end{tabular}}

%%%%%%%%%%%%%%%%%%%%%%%%%%%%%%%%%%%%%%%%%%
\begin{adjustwidth}{-\extralength}{0cm}

\reftitle{References}

\end{adjustwidth}

\begin{thebibliography}{999}

\bibitem[{Bernstein} {et~al.}(1982){Bernstein}, {Alexander}, {Arvidson},
  {Ely}, {Goad}, {Groth}, {Landgrebe}, {Legeckis}, {McPherron}, {Tananbaum},
  {Turrotte}, and {Vonder Haar}]{Bernstein1982}
{Bernstein}, R.; {Alexander}, S.S.; {Arvidson}, R.E.; {Ely}, O.; {Goad}, C.;
  {Groth}, E.J.; {Landgrebe}, D.A.; {Legeckis}, R.; {McPherron}, R.L.;
  {Tananbaum}, H.;  et~al.
\newblock {\em Data Management and Computation}; Volume 1: Issues and
  recommendations associated with distributed computation and data management
  systems for the space sciences; National Academy Press: Washington, DC, USA, 
  1982. https://doi.org/10.17226/19537.

\bibitem[{Arvidson} {et~al.}(1986){Arvidson}, {Landgrebe}, {Levinthal},
  {Ludwig}, {McCord}, {Schreier}, {Walker}, and {Wiederhold}]{Arvidson1986}
{Arvidson}, R.~E.and~Hunt, G.; {Landgrebe}, D.A.; {Levinthal}, E.; {Ludwig},
  G.H.; {McCord}, T.B.; {Schreier}, E.; {Walker}, R.; {Wiederhold}, G.
\newblock {\em Data Management and Computation}; Volume 2: Issues and
  recommendations associated with distributed computation and data management
  systems for the space sciences; National Academy Press: Washington, DC, USA,
  1986. https://doi.org/10.17226/12343.

\bibitem[PDS(2017)]{PDS2017}
Planetary Data System Roadmap Study for 2017--2026.
\newblock Technical Report, NASA Science Mission Directorate. 2017. Available online: 
  \url{https://pds.nasa.gov/home/about/PlanetaryDataSystemRMS17-26_20jun17.pdf} (accessed on 30 October 2022 %MDPI: Please add the access date (format: Date Month Year), e.g., accessed on 1 January 2020. %% NJC: done
).


\bibitem[{Data Design Working Group}(2021)]{PDS4concepts}
{Data Design Working Group}.
\newblock PDS4 Concepts.
\newblock Technical report, NASA PDS. 2021. Available online: 
  \url{https://pds.jpl.nasa.gov/datastandards/documents/concepts/Concepts_1.17.0.pdf} (accessed on 30 October 2022). %MDPI: Please add the access date (format: Date Month Year), e.g., accessed on 1 January 2020. %% NJC: done

\bibitem[{Prockter} {et~al.}(2021){Prockter}, {Tiscareno}, {Grayzeck},
  {Acton}, {Arvidson}, {Bauer}, {Beebe}, {Besse}, {Chanover}, {Crichton},
  {Gaddis}, {Gordon}, {Hare}, {Hollibaugh Baker}, {Hughes}, {Law}, {McAuley},
  {McClanahan}, {Padams}, {Raugh}, {Showalter}, {Stein}, and
  {Walker}]{Prockter2021}
{Prockter}, L.; {Tiscareno}, M.S.; {Grayzeck}, E.J.; {Acton}, C.H.; {Arvidson},
  R.E.; {Bauer}, J.M.; {Beebe}, R.; {Besse}, S.; {Chanover}, N.; {Crichton},
  D.J.;  et~al.
\newblock {The Planetary Data System: A Vital Component in NASA's Science
  Exploration Program}.
\newblock  \emph{Bull. Am. Astron. Soc.}  \textbf{2021}, \emph{53}, 459.
\newblock https://doi.org/10.3847/25c2cfeb.c3deebea.

\bibitem[Raugh and Hughes(2021)]{Raugh_2021}
Raugh, A.; Hughes, J.S.
\newblock The Road to an Archival Data Format{\textemdash}Data Structures.
\newblock {\em \psj} {\bf 2021}, {\em 2},~204.
\newblock https://doi.org/10.3847/\linebreak psj/ac1f22.

\bibitem[{Owen} and {Encrenaz}(2006)]{Owen2006}
{Owen}, T.; {Encrenaz}, T.
\newblock {Compositional constraints on giant planet formation}.
\newblock {\em \planss} {\bf 2006}, {\em 54},~1188--1196.
\newblock https://doi.org/10.1016/j.pss.2006.05.030.

\bibitem[{Gautier} {et~al.}(1981){Gautier}, {Conrath}, {Flasar}, {Hanel},
  {Kunde}, {Chedin}, and {Scott}]{Gautier1981}
{Gautier}, D.; {Conrath}, B.; {Flasar}, M.; {Hanel}, R.; {Kunde}, V.; {Chedin},
  A.; {Scott}, N.
\newblock {The helium abundance of Jupiter from Voyager}.
\newblock {\em \jgr} {\bf 1981}, {\em 86},~8713--8720.
\newblock https://doi.org/10.1029/JA086iA10p08713.

\bibitem[{Conrath} {et~al.}(1984){Conrath}, {Gautier}, {Hanel}, and
  {Hornstein}]{Conrath1984}
{Conrath}, B.J.; {Gautier}, D.; {Hanel}, R.A.; {Hornstein}, J.S.
\newblock {The helium abundance of Saturn from Voyager measurements}.
\newblock {\em \apj} {\bf 1984}, {\em 282},~807--815.
\newblock https://doi.org/10.1086/162267.

\bibitem[{Conrath} and {Gautier}(2000)]{Conrath2000}
{Conrath}, B.J.; {Gautier}, D.
\newblock {Saturn Helium Abundance: A Reanalysis of Voyager Measurements}.
\newblock {\em \icarus} {\bf 2000}, {\em 144},~124--134.
\newblock https://doi.org/10.1006/icar.1999.6265.

\bibitem[{Conrath} {et~al.}(1987){Conrath}, {Gautier}, {Hanel}, {Lindal},
  and {Marten}]{Conrath1987}
{Conrath}, B.; {Gautier}, D.; {Hanel}, R.; {Lindal}, G.; {Marten}, A.
\newblock {The helium abundance of Uranus from Voyager measurements}.
\newblock {\em \jgr} {\bf 1987}, {\em 92},~15003--15010.
\newblock  https://doi.org/10.1029/JA092iA13p15003.

\bibitem[{Conrath} {et~al.}(1991){Conrath}, {Gautier}, {Lindal},
  {Samuelson}, and {Shaffer}]{Conrath1991}
{Conrath}, B.J.; {Gautier}, D.; {Lindal}, G.F.; {Samuelson}, R.E.; {Shaffer},
  W.A.
\newblock {The helium abundance of Neptune from Voyager measurements}.
\newblock {\em \jgrs} {\bf 1991}, {\em 96},~18907.
\newblock https://doi.org/10.1029/91JA01703.

\bibitem[{Fouchet} {et~al.}(2009){Fouchet}, {Moses}, and
  {Conrath}]{Fouchet2009}
{Fouchet}, T.; {Moses}, J.I.; {Conrath}, B.J.
\newblock {Saturn: Composition and Chemistry}. In {\em Saturn from
  Cassini-Huygens}, Springer, New York, USA %MDPI: Please add the name of the publisher of it and their location (city country). %% NJC: Done
; {Dougherty}, M.K., {Esposito}, L.W., {Krimigis}, S.M.,
  Eds.;  2009; p.~83.
\newblock https://doi.org/10.1007/978-1-4020-9217-6\_5 .

\bibitem[{Fletcher} {et~al.}(2010){Fletcher}, {Achterberg}, {Greathouse},
  {Orton}, {Conrath}, {Simon-Miller}, {Teanby}, {Guerlet}, {Irwin}, and
  {Flasar}]{Fletcher2010}
{Fletcher}, L.N.; {Achterberg}, R.K.; {Greathouse}, T.K.; {Orton}, G.S.;
  {Conrath}, B.J.; {Simon-Miller}, A.A.; {Teanby}, N.; {Guerlet}, S.; {Irwin},
  P.G.J.; {Flasar}, F.M.
\newblock {Seasonal change on Saturn from Cassini/CIRS observations,
  2004--2009}.
\newblock {\em \icarus} {\bf 2010}, {\em 208},~337--352.
\newblock https://doi.org/10.1016/j.icarus.2010.01.022.

\bibitem[{Fletcher} {et~al.}(2018){Fletcher}, {Greathouse}, {Guerlet},
  {Moses}, and {West}]{Fletcher2018}
{Fletcher}, L.N.; {Greathouse}, T.K.; {Guerlet}, S.; {Moses}, J.I.; {West},
  R.A.
\newblock {Saturn's Seasonally Changing Atmosphere}. In {\emph{Saturn in the 21st
  Century}}, Cambridge University Press, Cambridge, England; {Baines}, K.H.; {Flasar}, F.M.; {Krupp}, N.; {Stallard}, T., Eds.;
  2018; pp. 251--294.
\newblock https://doi.org/10.1017/9781316227220.010.%MDPI: Please add the name of the publisher of it and their location (city country). %% NJC: Done

\bibitem[Smith {et~al.}(1979{\natexlab{a}})Smith, Soderblom, Johnson,
  Ingersoll, Collins, Shoemaker, Hunt, Masursky, Carr, Davies, Cook, Boyce,
  Danielson, Owen, Sagan, Beebe, Veverka, Strom, McCauley, Morrison, Briggs,
  and Suomi]{Smith1979a}
Smith, B.A.; Soderblom, L.A.; Johnson, T.V.; Ingersoll, A.P.; Collins, S.A.;
  Shoemaker, E.M.; Hunt, G.E.; Masursky, H.; Carr, M.H.; Davies, M.E.;  et~al.
\newblock The Jupiter System Through the Eyes of Voyager 1.
\newblock {\em Science} {\bf 1979}, {\em 204},~951--972.
\newblock https://doi.org/10.1126/science.204.4396.951.

\bibitem[Smith {et~al.}(1979{\natexlab{b}})Smith, Soderblom, Beebe, Boyce,
  Briggs, Carr, Collins, Cook, Danielson, Davies, Hunt, Ingersoll, Johnson,
  Masursky, McCauley, Morrison, Owen, Sagan, Shoemaker, Strom, Suomi, and
  Veverka]{Smith1979b}
Smith, B.A.; Soderblom, L.A.; Beebe, R.; Boyce, J.; Briggs, G.; Carr, M.;
  Collins, S.A.; Cook, A.F.; Danielson, G.E.; Davies, M.E.;  et~al.
\newblock The Galilean Satellites and Jupiter: Voyager 2 Imaging Science
  Results.
\newblock {\em Science} {\bf 1979}, {\em 206},~927--950.
\newblock https://doi.org/10.1126/science.206.4421.927.

\bibitem[Smith {et~al.}(1981)Smith, Soderblom, Beebe, Boyce, Briggs, Bunker,
Collins, Hansen, Johnson, Mitchell, Terrile, Carr, Cook, Cuzzi, Pollack,
  Danielson, Ingersoll, Davies, Hunt, Masursky, Shoemaker, Morrison, Owen,
  Sagan, Veverka, Strom, and Suomi]{Smith1981}
\textls[-15]{Smith, B.A.; Soderblom, L.; Beebe, R.; Boyce, J.; Briggs, G.; Bunker, A.;
  Collins, S.A.; Hansen, C.J.; Johnson, T.V.; Mitchell, J.L.;  et~al.
\newblock Encounter with Saturn: Voyager 1 Imaging Science Results.
\newblock {\em Science} {\bf 1981}, {\em 212},~163--191.
\newblock https://doi.org/10.1126/science.212.4491.163.}

\bibitem[Smith {et~al.}(1982)Smith, Soderblom, Batson, Bridges, Inge,
  Masursky, Shoemaker, Beebe, Boyce, Briggs, Bunker, Collins, Hansen, Johnson,
  Mitchell, Terrile, Cook, Cuzzi, Pollack, Danielson, Ingersoll, Davies, Hunt,
  Morrison, Owen, Sagan, Veverka, Strom, and Suomi]{Smith1982}
Smith, B.A.; Soderblom, L.; Batson, R.; Bridges, P.; Inge, J.; Masursky, H.;
  Shoemaker, E.; Beebe, R.; Boyce, J.; Briggs, G.;  et~al.
\newblock A New Look at the Saturn System: The Voyager 2 Images.
\newblock {\em Science} {\bf 1982}, {\em 215},~504--537.
\newblock https://doi.org/10.1126/science.215.4532.504.

\bibitem[Smith {et~al.}(1986)Smith, Soderblom, Beebe, Bliss, Boyce, Brahic,
  Briggs, Brown, Collins, Cook, Croft, Cuzzi, Danielson, Davies, Dowling,
  Godfrey, Hansen, Harris, Hunt, Ingersoll, Johnson, Krauss, Masursky,
  Morrison, Owen, Plescia, Pollack, Porco, Rages, Sagan, Shoemaker, Sromovsky,
  Stoker, Strom, Suomi, Synnott, Terrile, Thomas, Thompson, and
  Veverka]{Smith1986}
Smith, B.A.; Soderblom, L.A.; Beebe, R.; Bliss, D.; Boyce, J.M.; Brahic, A.;
  Briggs, G.A.; Brown, R.H.; Collins, S.A.; Cook, A.F.;  et~al.
\newblock Voyager 2 in the Uranian System: Imaging Science Results.
\newblock {\em Science} {\bf 1986}, {\em 233},~43--64.
\newblock https://doi.org/10.1126/science.233.4759.43.

\bibitem[Smith {et~al.}(1989)Smith, Soderblom, Banfield, c.~Barnet,
  Basilevsky, Beebe, Bollinger, Boyce, Brahic, Briggs, Brown, c.~Chyba,
  s.~A.~Collins, Colvin, Cook, Crisp, Croft, Cruikshank, Cuzzi, Danielson,
  Davies, Jong, Dones, Godfrey, Goguen, Grenier, Haemmerle, Hammel,
  c.~J.~Hansen, c.~P.~Helfenstein, Howell, Hunt, Ingersoll, Johnson, Kargel,
  Kirk, Kuehn, Limaye, Masursky, McEwen, Morrison, Owen, Owen, Pollack, c.~c.
  Porco, Rages, Rogers, Rudy, Sagan, Schwartz, Shoemaker, Showalter, Sicardy,
  Simonelli, Spencer, Sromovsky, Stoker, Strom, Suomi, Synott, Terrile, Thomas,
  Thompson, Verbiscer, and Veverka]{Smith1989}
Smith, B.A.; Soderblom, L.A.; Banfield, D.; c.~Barnet.; Basilevsky, A.T.;
  Beebe, R.F.; Bollinger, K.; Boyce, J.M.; Brahic, A.; Briggs, G.A.;  et~al.
\newblock Voyager 2 at Neptune: Imaging Science Results.
\newblock {\em Science} {\bf 1989}, {\em 246},~1422--1449.
\newblock https://doi.org/10.1126/science.246.4936.1422.

\bibitem[Vasavada {et~al.}(1998)Vasavada, Ingersoll, Banfield, Bell,
  Gierasch, Belton, Orton, Klaasen, DeJong, Breneman, Jones, Kaufman, Magee,
  and Senske]{Vasavada1998c}
Vasavada, A.R.; Ingersoll, A.P.; Banfield, D.; Bell, M.; Gierasch, P.J.;
  Belton, M.J.; Orton, G.S.; Klaasen, K.P.; DeJong, E.; Breneman, H.;  et~al.
\newblock Galileo Imaging of Jupiter's Atmosphere: The Great Red Spot,
  Equatorial Region, and White Ovals.
\newblock {\em Icarus} {\bf 1998}, {\em 135},~265--275.
\newblock https://doi.org/10.1006/icar.1998.5984.

\bibitem[Porco {et~al.}(2005)Porco, Baker, Barbara, Beurle, Brahic, Burns,
  Charnoz, Cooper, Dawson, Genio, Denk, Dones, Dyudina, Evans, Giese, Grazier,
  Helfenstein, Ingersoll, Jacobson, Johnson, McEwen, Murray, Neukum, Owen,
  Perry, Roatsch, Spitale, Squyres, Thomas, Tiscareno, Turtle, Vasavada,
  Veverka, Wagner, and West]{Porco2005c}
Porco, C.C.; Baker, E.; Barbara, J.; Beurle, K.; Brahic, A.; Burns, J.A.;
  Charnoz, S.; Cooper, N.; Dawson, D.D.; Genio, A.D.D.;  et~al.
\newblock Cassini Imaging Science: Initial Results on Saturn's Atmosphere.
\newblock {\em Science} {\bf 2005}, {\em 307},~1243--1247.
\newblock https://doi.org/10.1126/science.1107691.

\bibitem[García-Melendo {et~al.}(2011)García-Melendo, Pérez-Hoyos,
  Sánchez-Lavega, and Hueso]{Garcia-Melendo2011a}
García-Melendo, E.; Pérez-Hoyos, S.; Sánchez-Lavega, A.; Hueso, R.
\newblock Saturn’s zonal wind profile in 2004–2009 from Cassini ISS images
  and its long-term variability.
\newblock {\em Icarus} {\bf 2011}, {\em 215},~62--74.
\newblock https://doi.org/10.1016/j.icarus.2011.07.005.

\bibitem[Porco {et~al.}(2003)Porco, West, McEwen, Genio, Ingersoll, Thomas,
  Squyres, Dones, Murray, Johnson, Burns, Brahic, Neukum, Veverka, Barbara,
  Denk, Evans, Ferrier, Geissler, Helfenstein, Roatsch, Throop, Tiscareno, and
  Vasavada]{Porco2003a}
Porco, C.C.; West, R.A.; McEwen, A.; Genio, A.D.D.; Ingersoll, A.P.; Thomas,
  P.; Squyres, S.; Dones, L.; Murray, C.D.; Johnson, T.V.;  et~al.
\newblock Cassini Imaging of Jupiter's Atmosphere, Satellites, and Rings.
\newblock {\em Science} {\bf 2003}, {\em 299},~1541--1547.
\newblock https://doi.org/10.1126/science.1079462.

\bibitem[Li {et~al.}(2006)Li, Ingersoll, Vasavada, Simon-Miller, Genio,
  Ewald, Porco, and West]{Li2006a}
Li, L.; Ingersoll, A.P.; Vasavada, A.R.; Simon-Miller, A.A.; Genio, A.D.D.;
  Ewald, S.P.; Porco, C.C.; West, R.A.
\newblock Vertical wind shear on Jupiter from Cassini images.
\newblock {\em \jgr} {\bf 2006}, {\em 111},~E04004.
\newblock https://doi.org/10.1029/2005JE002556.

\bibitem[Simon {et~al.}(2015)Simon, Li, and Reuter]{Simon2015}
Simon, A.A.; Li, L.; Reuter, D.C.
\newblock Small-scale waves on Jupiter: A reanalysis of New Horizons, Voyager,
  and Galileo data.
\newblock {\em \grl} {\bf 2015}, {\em 42},~2612--2618.
\newblock https://doi.org/10.1002/2015GL063433.

\bibitem[{Simon} {et~al.}(2014){Simon}, {Wong}, {Rogers}, {Orton}, {de
  Pater}, {Asay-Davis}, {Carlson}, and {Marcus}]{Simon2014}
{Simon}, A.A.; {Wong}, M.H.; {Rogers}, J.H.; {Orton}, G.S.; {de Pater}, I.;
  {Asay-Davis}, X.; {Carlson}, R.W.; {Marcus}, P.S.
\newblock {Dramatic Change in Jupiter's Great Red Spot from Spacecraft
  Observations}.
\newblock {\em \apjl} {\bf 2014}, {\em 797},~L31.
\newblock https://doi.org/10.1088/2041-8205/797/2/L31 .

\bibitem[{Sanchez-Lavega} {et~al.}(2018){Sanchez-Lavega}, {Fischer},
  {Fletcher}, {Garcia-Melendo}, {Hesman}, {Perez-Hoyos}, {Sayanagi}, and
  {Sromovsky}]{SanchezLavega2018}
{Sanchez-Lavega}, A.; {Fischer}, G.; {Fletcher}, L.N.; {Garcia-Melendo}, E.;
  {Hesman}, B.; {Perez-Hoyos}, S.; {Sayanagi}, K.M.; {Sromovsky}, L.A.
\newblock {The Great Saturn Storm of 2010{\textendash}2011}. In {\emph{Saturn in
  the 21st Century}}, Cambridge University Press, Cambridge, England; {Baines}, K.H., {Flasar}, F.M., {Krupp}, N., {Stallard},
  T., Eds.;  2018; pp. 377--416.
\newblock https://doi.org/10.1017/9781316227220.013.%MDPI: Please add the name of the publisher of it and their location (city country). %% NJC: Done

\bibitem[{Elliot} and {Olkin}(1996)]{Elliot1996}
{Elliot}, J.L.; {Olkin}, C.B.
\newblock {Probing Planetary Atmospheres with Stellar Occultations}.
\newblock {\em \areps} {\bf 1996}, {\em 24},~89--124.
\newblock https://doi.org/10.1146/annurev.earth.24.1.89.

\bibitem[{Greathouse} {et~al.}(2010){Greathouse}, {Gladstone}, {Moses},
  {Stern}, {Retherford}, {Vervack}, {Slater}, {Versteeg}, {Davis}, {Young},
  {Steffl}, {Throop}, and {Parker}]{Greathouse2010}
{Greathouse}, T.K.; {Gladstone}, G.R.; {Moses}, J.I.; {Stern}, S.A.;
  {Retherford}, K.D.; {Vervack}, R.J.; {Slater}, D.C.; {Versteeg}, M.H.;
  {Davis}, M.W.; {Young}, L.A.;  et~al.
\newblock {New Horizons Alice ultraviolet observations of a stellar occultation
  by Jupiter{\textquoteright}s atmosphere}.
\newblock {\em \icarus} {\bf 2010}, {\em 208},~293--305.
\newblock https://doi.org/10.1016/j.icarus.2010.02.002 .

\bibitem[{Fletcher} {et~al.}(2020){Fletcher}, {Orton}, {Greathouse},
  {Rogers}, {Zhang}, {Oyafuso}, {Eichst{\"a}dt}, {Melin}, {Li}, {Levin},
  {Bolton}, {Janssen}, {Mettig}, {Grassi}, {Mura}, and {Adriani}]{Fletcher2020}
{Fletcher}, L.N.; {Orton}, G.S.; {Greathouse}, T.K.; {Rogers}, J.H.; {Zhang},
  Z.; {Oyafuso}, F.A.; {Eichst{\"a}dt}, G.; {Melin}, H.; {Li}, C.; {Levin},
  S.M.;  et~al.
\newblock {Jupiter's Equatorial Plumes and Hot Spots: Spectral Mapping from
  Gemini/TEXES and Juno/MWR}.
\newblock {\em J. Geophys. Res. (Planets)} {\bf 2020}, {\em
  125},~e06399.
\newblock https://doi.org/10.1029/2020JE006399.

\bibitem[{Atkinson} {et~al.}(1997){Atkinson}, {Ingersoll}, and
  {Seiff}]{Atkinson1997}
{Atkinson}, D.H.; {Ingersoll}, A.P.; {Seiff}, A.
\newblock {Deep winds on Jupiter as measured by the Galileo probe}.
\newblock {\em Nature} {\bf 1997}, {\em 388},~649--650.
\newblock https://doi.org/10.1038/41718.

\bibitem[{Niemann} {et~al.}(1998){Niemann}, {Atreya}, {Carignan}, {Donahue},
  {Haberman}, {Harpold}, {Hartle}, {Hunten}, {Kasprzak}, {Mahaffy}, {Owen}, and
  {Way}]{Niemann1998}
{Niemann}, H.B.; {Atreya}, S.K.; {Carignan}, G.R.; {Donahue}, T.M.; {Haberman},
  J.A.; {Harpold}, D.N.; {Hartle}, R.E.; {Hunten}, D.M.; {Kasprzak}, W.T.;
  {Mahaffy}, P.R.;  et~al.
\newblock {The composition of the Jovian atmosphere as determined by the
  Galileo probe mass spectrometer}.
\newblock {\em \jgr} {\bf 1998}, {\em 103},~22831--22846.
\newblock https://doi.org/10.1029/98JE01050.

\bibitem[{Rinnert} {et~al.}(1998){Rinnert}, {Lanzerotti}, {Uman}, {Dehmel},
  {Gliem}, {Krider}, and {Bach}]{Rinnert1998}
{Rinnert}, K.; {Lanzerotti}, L.J.; {Uman}, M.A.; {Dehmel}, G.; {Gliem}, F.O.;
  {Krider}, E.P.; {Bach}, J.
\newblock {Measurements of radio frequency signals from lightning in Jupiter's
  atmosphere}.
\newblock {\em \jgr} {\bf 1998}, {\em 103},~22979--22992.
\newblock https://doi.org/10.1029/98JE00830.

\bibitem[{Nagy} {et~al.}(2009){Nagy}, {Kliore}, {Mendillo}, {Miller},
  {Moore}, {Moses}, {M{\"u}ller-Wodarg}, and {Shemansky}]{Nagy2009}
{Nagy}, A.F.; {Kliore}, A.J.; {Mendillo}, M.; {Miller}, S.; {Moore}, L.;
  {Moses}, J.I.; {M{\"u}ller-Wodarg}, I.; {Shemansky}, D.
\newblock {Upper Atmosphere and Ionosphere of Saturn}. In {\em Saturn from
  Cassini-Huygens}, Springer, New York, USA; {Dougherty}, M.K., {Esposito}, L.W., {Krimigis}, S.M.,
  Eds.;  2009; p. 181.
\newblock https://doi.org/10.1007/978-1-4020-9217-6\_8.%MDPI: Please add the name of the publisher of it and their location (city country). %% NJC: Done

\bibitem[{Bader} {et~al.}(2020){Bader}, {Cowley}, {Badman}, {Ray},
  {Kinrade}, {Palmaerts}, and {Pryor}]{Bader2020}
{Bader}, A.; {Cowley}, S.W.H.; {Badman}, S.V.; {Ray}, L.C.; {Kinrade}, J.;
  {Palmaerts}, B.; {Pryor}, W.R.
\newblock {The Morphology of Saturn's Aurorae Observed During the Cassini Grand
  Finale}.
\newblock {\em \grl} {\bf 2020}, {\em 47},~e85800.
\newblock https://doi.org/10.1029/2019GL085800.

\bibitem[{Dinelli} {et~al.}(2019){Dinelli}, {Adriani}, {Mura}, {Altieri},
  {Migliorini}, and {Moriconi}]{Dinelli2019}
{Dinelli}, B.M.; {Adriani}, A.; {Mura}, A.; {Altieri}, F.; {Migliorini}, A.;
  {Moriconi}, M.L.
\newblock {JUNO/JIRAM's view of Jupiter's H$_{3}^{+}$ emissions}.
\newblock {\em Philos. Trans. R. Soc. Lond. Ser.
  A} {\bf 2019}, {\em 377},~20180406.
\newblock https://doi.org/10.1098/rsta.2018.0406.

\bibitem[{Lockwood}(2019{\natexlab{a}})]{Lockwood2019b}
{Lockwood}, G.W.
\newblock {Final compilation of photometry of Uranus and Neptune, 1972--2016.}
\newblock {\em Icarus} {\bf 2019}, {\em 324},~77--85.
\newblock https://doi.org/10.1016/j.icarus.2019.01.024.

\bibitem[{Lockwood}(2019{\natexlab{b}})]{Lockwood2019a}
{Lockwood}, G.W.
\newblock Lowell Observatory Uranus and Neptune b (472 nm) and y (551 nm)
  Photometry (1972--2016) Data Archive {\bf 2019}.
\newblock PDS Atmospheres Node (ATM). \newblock https://doi.org/10.17189/1518957.
%MDPI: Please provide more information about the article type, such as book (please provide the name and location of the publisher); online resource (please provide the URL of the website and the date it was accessed (Date Month Year)); or journal article (please provide the name of the journal, the year and volume in which it was published, and the page number). Please refer to https://www.mdpi.com/authors/references for full reference formatting guides.
%% NJC: This is an online data citation, not a web site. The data themselves have a permanent DOI just like a journal paper. The formatting rules for citing such data are given here: https://pds.nasa.gov/datastandards/citing/  We modified the citation accordingly.

\bibitem[{Burke} and {Franklin}(1955)]{burke1955}
{Burke}, B.F.; {Franklin}, K.L.
\newblock {Observations of a Variable Radio Source Associated with the Planet
  Jupiter}.
\newblock {\em J. Geophys. Res.} {\bf 1955}, {\em 60},~213--217.
\newblock https://doi.org/10.1029/JZ060i002p00213.

\bibitem[{Gurnett} {et~al.}(2007){Gurnett}, {Persoon}, {Kurth}, {Groene},
  {Averkamp}, {Dougherty}, and {Southwood}]{Gurnett2007}
{Gurnett}, D.; {Persoon}, A.; {Kurth}, W.; {Groene}, J.; {Averkamp}, T.;
  {Dougherty}, M.; {Southwood}, D.
\newblock {The Variable Rotation Peripod of the Inner Region of Saturn’s
  Plasma Disk}.
\newblock {\em Science} {\bf 2007}, {\em 316},~442--445.
\newblock https://doi.org/doi:10.1126/science.1138562.

\bibitem[{Zarka}(1998)]{zarka1998}
{Zarka}, P.
\newblock {Auroral radio emissions at the outer planets: Observations and
  theories}.
\newblock {\em J. Geophys. Res.} {\bf 1998}, {\em 103},~20159--20194.
\newblock https://doi.org/10.1029/98JE01323.

\bibitem[{Kaspi} {et~al.}(2020){Kaspi}, {Galanti}, {Showman}, {Stevenson},
  {Guillot}, {Iess}, and {Bolton}]{Kaspi2020}
{Kaspi}, Y.; {Galanti}, E.; {Showman}, A.P.; {Stevenson}, D.J.; {Guillot}, T.;
  {Iess}, L.; {Bolton}, S.J.
\newblock {Comparison of the Deep Atmospheric Dynamics of Jupiter and Saturn in
  Light of the Juno and Cassini Gravity Measurements}.
\newblock {\em \ssr} {\bf 2020}, {\em 216},~84.
\newblock https://doi.org/10.1007/s11214-020-00705-7.

\bibitem[{Iess} {et~al.}(2018){Iess}, {Folkner}, {Durante}, {Parisi},
  {Kaspi}, {Galanti}, {Guillot}, {Hubbard}, {Stevenson}, {Anderson}, {Buccino},
  {Casajus}, {Milani}, {Park}, {Racioppa}, {Serra}, {Tortora}, {Zannoni},
  {Cao}, {Helled}, {Lunine}, {Miguel}, {Militzer}, {Wahl}, {Connerney},
  {Levin}, and {Bolton}]{Iess2018}
{Iess}, L.; {Folkner}, W.M.; {Durante}, D.; {Parisi}, M.; {Kaspi}, Y.;
  {Galanti}, E.; {Guillot}, T.; {Hubbard}, W.B.; {Stevenson}, D.J.; {Anderson},
  J.D.;  et~al.
\newblock {Measurement of Jupiter{\textquoteright}s asymmetric gravity field}.
\newblock {\em \nat} {\bf 2018}, {\em 555},~220--222.
\newblock https://doi.org/10.1038/nature25776.

\bibitem[{Hedman} {et~al.}(2022){Hedman}, {Nicholson}, {El Moutamid}, and
  {Smotherman}]{Hedman22}
{Hedman}, M.M.; {Nicholson}, P.D.; {El Moutamid}, M.; {Smotherman}, S.
\newblock {Kronoseismology. VI. Reading the recent history of Saturn's gravity
  field in its rings}.
\newblock {\em \psj} {\bf 2022}, {\em 3},~61. 
\newblock https://doi.org/10.3847/PSJ/ac4df8.

\bibitem[{Galilei}(1610)]{Galilei1610}
{Galilei}, G.
\newblock {\em Sidereus Nuncius Magna, Longeque Admirabilia Spectacula Pandens
  Lunae Facie, Fixis Innumeris, Lacteo Circulo, Stellis Nebulosis, $\ldots$ Galileo
  Galileo: Nuper a se Reperti Beneficio Sunt Observata in Apprime vero in
  Quatuor Planetis Circa Iovis Stellam Disparibus Intervallis, Atque Periodis,
  Celeritate Mirabili Circumvolutis $\ldots$ atque Medicea Sidera Nuncupandos %MDPI: %MDPI: Please add the name of the publisher of it and their location (city country).. %% NJC: Done
  Decrevit}; Thomas Baglioni, Republic of Venice, {\bf 1610}.
\newblock https://doi.org/10.3931/e-rara-695.

\bibitem[Hastie {et~al.}(2009)Hastie, Tibshirani, Friedman, and
  Friedman]{hastie2009elements}
Hastie, T.; Tibshirani, R.; Friedman, J.H.; Friedman, J.H.
\newblock {\em The Elements of Statistical Learning: Data Mining, Inference,
  Furthermore, Prediction}; Springer: Berlin/Heidelberg, Germany, %newly added information, please confirm  %% NJC: looks fine
  2009; Volume~2, 

\bibitem[{Edgington} {et~al.}(2019){Edgington}, {Tapella}, {Beebe},
  {Buratti}, {Burton}, {Weld}, {Streiffert}, {Connell}, {Ray}, {Brooks}, and
  {Evans}]{Edgington2019}
{Edgington}, S.G.; {Tapella}, R.K.; {Beebe}, R.F.; {Buratti}, B.J.; {Burton},
  M.E.; {Weld}, K.R.; {Streiffert}, B.A.; {Connell}, A.M.; {Ray}, T.L.;
  {Brooks}, S.M.;  et~al.
\newblock {Cassini-Huygens Scientific Legacy: The Cassini Mission Archive at
  the Planetary Data System}.
\newblock   In Proceedings of the 50th Annual Lunar and Planetary Science Conference, 2019 %MDPI: Please add the location (city country) and date (day month year) of the conference. %% NJC: Done 
; The Woodlands, TX, USA, 18--22 March 2019; Lunar and Planetary
  Institute: Houston, TX, USA, 2019; p. 2932.

\bibitem[Bailen {et~al.}(2013)Bailen, Sucharski, Akins, Hare, and
  Gaddis]{Bailen:lpsc2013}
Bailen, M.S.; Sucharski, R.M.; Akins, S.W.; Hare, T.M.; Gaddis, L.R.
\newblock Using the PDS Planetary Image Locator Tool (PILOT) to Identify and
  Download Spacecraft Data for Research.
\newblock  In Proceedings of the 44th Lunar and Planetary Science Conference, The Woodlands, TX, USA, 18--22 March %MDPI: Newly added information. Please confirm. %% NJC: Done
 2013; Lunar and Planetary
  Institute: Houston, TX, USA, 2013; p. Abstract \#2246.

\bibitem[Akins {et~al.}(2009)Akins, Gaddis, Becker, Barrett, Bailen, Hare,
  Soderblom, and Raub]{Akins:lpsc2009}
Akins, S.W.; Gaddis, L.; Becker, K.; Barrett, J.; Bailen, M.; Hare, T.;
  Soderblom, L.A.; Raub, R.
\newblock Status of the PDS Unified Planetary Coordinates Database and the
  Planetary Image Locator Tool (PILOT).
\newblock  In Proceedings of the 40th Lunar and Planetary Science Conference, The Woodlands, TX, USA, 23--27 March 2009; Lunar and Planetary
  Institute: Houston, TX, USA, 2009; p. Abstract \#2002. %MDPI: Newly added information. Please confirm. %% Done

\bibitem[Hare {et~al.}(2013)Hare, Akins, Sucharski, Bailen, and
  Anderson]{Hare:lpsc2013}
Hare, T.M.; Akins, S.W.; Sucharski, R.M.; Bailen, M.S.; Anderson, J.A.
\newblock Map Projection Web Service for PDS Images.
\newblock  In Proceedings of the 44th Lunar and Planetary Science Conference,  The Woodlands, TX, USA, 18--22 March 2013; Lunar and Planetary
  Institute: Houston, TX, USA, 2013; p. Abstract \#2068. %MDPI: Newly added information. Please confirm. %% NJC: Done

\bibitem[Akins {et~al.}(2015)Akins, Hare, Sucharski, Bailen, and
  Gaddis]{Akins:planetdata2015}
Akins, S.W.; Hare, T.M.; Sucharski, R.M.; Bailen, M.S.; Gaddis, L.R.
\newblock POW and MAP2: Job Management and Advanced Processing.
\newblock  In Proceedings of the Second Planetary Data Workshop, Flagstaff, AZ, USA, 8--11 June 2015; Lunar and Planetary Institute:
  Houston, TX, USA, 2015; p. Abstract \#7037. %MDPI: Newly added information. Please confirm. %% NJC: Done

\bibitem[Sides {et~al.}(2017)Sides, Becker, Becker, Edmundson, Backer,
  Wilson, Weller, Humphrey, Berry, Shepherd, Hahn, Rose, Rodriguez, Paquette,
  Mapel, Shinaman, and Richie]{Sides:lpsc2017}
Sides, S.C.; Becker, T.L.; Becker, K.J.; Edmundson, K.L.; Backer, J.W.; Wilson,
  T.J.; Weller, L.A.; Humphrey, I.R.; Berry, K.L.; Shepherd, M.R.;  et~al.
\newblock The USGS Integrated Software for Imagers and Spectrometers (ISIS 3)
  Instrument Support, New Capabilities, and Releases.
\newblock  In Proceedings of the 48th Lunar and Planetary Science Conference, The Woodlands, TX, USA, 20--24 March 2017; Lunar and Planetary 
  Institute: Houston, TX, USA, 2017; p. Abstract \#2739. %MDPI: Newly added information. Please confirm. %% NJC: Done

\bibitem[Hare and Plesea(2008)]{Hare:lpsc2008}
Hare, T.M.; Plesea, L.
\newblock Planetary GIS Updates for 2007.
\newblock  In Proceedings of the 39th Lunar and Planetary Science Conference; Lunar and Planetary
  Institute: Houston, TX, USA, 2008; p. Abstract \#2536. %MDPI: Newly added information. Please confirm. %% NJC: Done

\bibitem[Padams {et~al.}(2017)Padams, Grimes, Hollins, Lavoie, and
  Stanboli]{Padams:planetdata2017}
Padams, J.; Grimes, K.; Hollins, G.; Lavoie, S.; Stanboli, A.
\newblock NASA PDS IMG: Accessing Your Planetary Image Data.
\newblock  In Proceedings of the 3rd Planetary Data Workshop, Flagstaff, AZ, USA, 12--15 June 2017; Lunar and Planetary Institute: Houston, TX, USA, 
   2017; p. Abstract \#7114. %MDPI: Newly added information. Please confirm. %% NJC: Done

\bibitem[Altinok {et~al.}(2014)Altinok, Bornstein, Estlin, Gaines, Schaffer,
  Thompson, Anderson, Burl, Castano, and Wiens]{Altinok:lpsc2014}
Altinok, A.; Bornstein, B.; Estlin, T.; Gaines, D.; Schaffer, S.; Thompson,
  D.R.; Anderson, R.C.; Burl, M.; Castano, R.; Wiens, R.
\newblock Automatic Image Analysis for Adaptive Instrument Targeting:
  Applications to MSL and Mars 2020.
\newblock  In Proceedings of the 45th Lunar and Planetary Science Conference, The Woodlands, TX, USA, 17--21 March 2014; Lunar and Planetary
  Institute: Houston, TX, USA, 2014; p. Abstract \#2871.

\bibitem[Stanboli {et~al.}(2017)Stanboli, Bue, Wagstaff, and
  Altinok]{Stanboli:planetdata2017}
Stanboli, A.; Bue, B.; Wagstaff, K.; Altinok, A.
\newblock Automated Content Detection for Cassini Images.
\newblock  In Proceedings of the 3rd Planetary Data Workshop, Flagstaff, AZ, USA, 12--15 June 2017; Lunar and Planetary Institute: Houston, TX, USA,
   2017; \linebreak p. Abstract \#7048.

\bibitem[{Garc\'{i}a-Melendo} and
  {S\'{a}nchez-Lavega}(2011)]{GarciaMelendo2011}
{Garc\'{i}a-Melendo}, E.; {S\'{a}nchez-Lavega}, A.
\newblock Saturn Zonal Wind Bundle Data Archive {\bf 2011}.
\newblock PDS Atmospheres Node (ATM). \newblock {https://doi.org/10.17189/1518962}.
%MDPI: Please provide more information about the article type, such as book (please provide the name and location of the publisher); online resource (please provide the URL of the website and the date it was accessed (Date Month Year)); or journal article (please provide the name of the journal, the year and volume in which it was published, and the page number). Please refer to https://www.mdpi.com/authors/references for full reference formatting guides.
%% NJC: This is an online data citation, not a web site. The data themselves have a permanent DOI just like a journal paper. The formatting rules for citing such data are given here: https://pds.nasa.gov/datastandards/citing/  We modified the citation accordingly.

\bibitem[{Koskinen}(2018)]{Koskinen2018}
{Koskinen}, T.
\newblock Structure of Saturn's Thermosphere from Stellar Occultations Bundle Data Archive {\bf 2018}.
\newblock PDS Atmospheres Node (ATM). \newblock {https://doi.org/10.17189/518e-p721.}
%MDPI: Please provide more information about the article type, such as book (please provide the name and location of the publisher); online resource (please provide the URL of the website and the date it was accessed (Date Month Year)); or journal article (please provide the name of the journal, the year and volume in which it was published, and the page number). Please refer to https://www.mdpi.com/authors/references for full reference formatting guides.
%% NJC: This is an online data citation, not a web site. The data themselves have a permanent DOI just like a journal paper. The formatting rules for citing such data are given here: https://pds.nasa.gov/datastandards/citing/  We modified the citation accordingly.

\bibitem[{Bernath}(2022)]{Bernath2022}
{Bernath}, P.
\newblock Laboratory Study of Hydrocarbon IR Spectra Bundle Data Archive {\bf 2022}.
newblock PDS Atmospheres Node (ATM). \newblock {https://doi.org/10.17189/1518949}.
%MDPI: Please provide more information about the article type, such as book (please provide the name and location of the publisher); online resource (please provide the URL of the website and the date it was accessed (Date Month Year)); or journal article (please provide the name of the journal, the year and volume in which it was published, and the page number). Please refer to https://www.mdpi.com/authors/references for full reference formatting guides.
%% NJC: This is an online data citation, not a web site. The data themselves have a permanent DOI just like a journal paper. The formatting rules for citing such data are given here: https://pds.nasa.gov/datastandards/citing/  We modified the citation accordingly.

\bibitem[{Newman} {et~al.}(2021){Newman}, {Airapetian}, {Battalio},
  {Bougher}, {Brown}, {Domagal-Goldman}, {Fan}, {Guzewich}, {Heavens},
  {Jackson}, {Kahre}, {Mischna}, {McConnochie}, {Neakrase}, {Pankine},
  {Pla-Garc{\'\i}a}, {Richardson}, {Smith}, {Solomonidou}, {Soto}, {Toigo}, and
  {Vi{\'u}dez-Moreiras}]{Newman2021}
{Newman}, C.; {Airapetian}, V.; {Battalio}, M.; {Bougher}, S.; {Brown}, A.;
  {Domagal-Goldman}, S.D.; {Fan}, S.; {Guzewich}, S.D.; {Heavens}, N.G.;
  {Jackson}, D.;  et~al.
\newblock {An Urgently Needed Repository for Planetary Atmospheric Model
  Output}.
\newblock  \emph{Bull. Am. Astron. Soc.} \textbf{2021}, \emph{53}, 
  505. https://doi.org/10.3847/25c2cfeb.6974fd2e.

\bibitem[PDE(2021)]{PDEIRB2021}
Final Report of the Planetary Data Ecosystem Independent Review Board.
\newblock Technical Report, NASA Science Mission Directorate. 2021.  Available online: 
  \url{https://science.nasa.gov/science-pink/s3fs-public/atoms/files/PDE\%20IRB\%20Final\%20Report.pdf}  (accessed on 30 October 2022).%MDPI: Please add accessed on date %% NJC: Done

\bibitem[{Hsu} {et~al.}(2019){Hsu}, {Wong}, and {Simon}]{Hsu2019}
{Hsu}, A.I.; {Wong}, M.H.; {Simon}, A.A.
\newblock {Lifetimes and Occurrence Rates of Dark Vortices on Neptune from 25
  Years of Hubble Space Telescope Images}.
\newblock {\em \aj} {\bf 2019}, {\em 157},~152. https://doi.org/10.3847/1538-3881/ab0747.

\bibitem[{Clarke} {et~al.}(2005){Clarke}, {G{\'e}rard}, {Grodent},
  {Wannawichian}, {Gustin}, {Connerney}, {Crary}, {Dougherty}, {Kurth},
  {Cowley}, {Bunce}, {Hill}, and {Kim}]{Clarke2005}
{Clarke}, J.T.; {G{\'e}rard}, J.C.; {Grodent}, D.; {Wannawichian}, S.;
  {Gustin}, J.; {Connerney}, J.; {Crary}, F.; {Dougherty}, M.; {Kurth}, W.;
  {Cowley}, S.W.H.;  et~al.
\newblock {Morphological differences between Saturn's ultraviolet aurorae and
  those of Earth and Jupiter}.
\newblock {\em \nat} {\bf 2005}, {\em 433},~717--719. https://doi.org/10.1038/nature03331.

\bibitem[{Tollefson} {et~al.}(2017){Tollefson}, {Wong}, {Pater}, {Simon},
  {Orton}, {Rogers}, {Atreya}, {Cosentino}, {Januszewski},
  {Morales-Juber{\'\i}as}, and {Marcus}]{Tollefson2017}
{Tollefson}, J.; {Wong}, M.H.; {Pater}, I.d.; {Simon}, A.A.; {Orton}, G.S.;
  {Rogers}, J.H.; {Atreya}, S.K.; {Cosentino}, R.G.; {Januszewski}, W.;
  {Morales-Juber{\'\i}as}, R.;  et~al.
\newblock {Changes in Jupiter's Zonal Wind Profile preceding and during the
  Juno mission}.
\newblock {\em \icarus} {\bf 2017}, {\em 296},~163--178. https://doi.org/10.1016/j.icarus.2017.06.007.

\bibitem[{Karkoschka}(1994)]{Karkoschka1994}
{Karkoschka}, E.
\newblock {Spectrophotometry of the Jovian Planets and Titan at 300- to 1000-nm
  Wavelength: The Methane Spectrum}.
\newblock {\em \icarus} {\bf 1994}, {\em 111},~174--192.
\newblock https://doi.org/10.1006/icar.1994.1139.

\bibitem[{Karkoschka}(1998)]{Karkoschka1998}
{Karkoschka}, E.
\newblock Spectrophotometry of the Jovian Planets and Titan Data Archive {\bf 1998}.
newblock PDS Atmospheres Node (ATM). \newblock https://doi.org/10.17189/2bp8-k793.
%MDPI: Please provide more information about the article type, such as book (please provide the name and location of the publisher); online resource (please provide the URL of the website and the date it was accessed (Date Month Year)); or journal article (please provide the name of the journal, the year and volume in which it was published, and the page number). Please refer to https://www.mdpi.com/authors/references for full reference formatting guides.
%% NJC: This is an online data citation, not a web site. The data themselves have a permanent DOI just like a journal paper. The formatting rules for citing such data are given here: https://pds.nasa.gov/datastandards/citing/  We modified the citation accordingly.

\bibitem[{Mayorga} {et~al.}(2016){Mayorga}, {Jackiewicz}, {Rages}, {West},
  {Knowles}, {Lewis}, and {Marley}]{Mayorga2016}
{Mayorga}, L.C.; {Jackiewicz}, J.; {Rages}, K.; {West}, R.A.; {Knowles}, B.;
  {Lewis}, N.; {Marley}, M.S.
\newblock {Jupiter{\textquoteright}s Phase Variations from Cassini: A Testbed
  for Future Direct-imaging Missions}.
\newblock {\em \aj} {\bf 2016}, {\em 152},~209. https://doi.org/10.3847/0004-6256/152/6/209.

\bibitem[{Gupta} {et~al.}(2022){Gupta}, {Atreya}, {Steffes}, {Fletcher},
  {Guillot}, {Allison}, {Bolton}, {Helled}, {Levin}, {Li}, {Lunine}, {Miguel},
  {Orton}, {Hunter Waite}, and {Withers}]{Gupta2022}
{Gupta}, P.; {Atreya}, S.K.; {Steffes}, P.G.; {Fletcher}, L.N.; {Guillot}, T.;
  {Allison}, M.D.; {Bolton}, S.J.; {Helled}, R.; {Levin}, S.; {Li}, C.;  et~al.
\newblock {Jupiter's Temperature Structure: A Reassessment of the Voyager Radio
  Occultation Measurements}.
\newblock {\em \psj} {\bf 2022}, {\em 3},~159. https://doi.org/10.3847/PSJ/ac6956.

\end{thebibliography}
\end{document}